\RequirePackage[2014/01/01]{latexrelease}
\documentclass[journal]{IEEEtran}

\ifCLASSINFOpdf
\else
\fi
\usepackage{amsmath,amstext,amsfonts,amssymb,amsthm,latexsym}
\usepackage{times,cite}
\usepackage[T1]{fontenc}
\usepackage[utf8]{inputenc}
\usepackage{graphicx}
\usepackage{algpseudocode}
\usepackage{caption}
\usepackage{newfloat}
\usepackage[table,rgb]{xcolor}
\usepackage{lipsum}
\usepackage{subcaption}
\usepackage{multirow}
\usepackage{tikz}
\usepackage{verbatim}
\usepackage{makecell}
\usepackage{ifthen}

\usetikzlibrary{fit,calc,positioning,decorations.pathreplacing,matrix}
\usetikzlibrary{calc}

\newcommand{\norm}[1]{\left\lVert#1\right\rVert}
\newcommand{\argmin}{\operatornamewithlimits{argmin}}
\newcommand{\prox}{\text{prox}}

\newcommand{\hl}[1]{#1}

\DeclareFloatingEnvironment[fileext=frm,placement={!ht},name=Algorithm]{algorithm}
\begin{document}
\IEEEoverridecommandlockouts
\IEEEpubid{\begin{minipage}[t]{\textwidth}\ \\[10pt]
        \centering\small{Copyright \copyright 2019 IEEE. Personal use of this material is permitted. However, permission to use this material for any other purposes must be obtained from the IEEE by sending a request to pubs-permissions@ieee.org.}
\end{minipage}}

\title{A Partially Learned Algorithm for Joint Photoacoustic Reconstruction and Segmentation}

\author{Yoeri~E.~Boink, Srirang~Manohar~and~Christoph~Brune\vspace{-5mm}%
\thanks{Y.E. Boink and C. Brune are with the Department
of Applied Mathematics, University of Twente, Enschede, 7500 AE, Netherlands. e-mail: y.e.boink@utwente.nl}%
\thanks{Y.E. Boink and S. Manohar are with the Biomedical Photonic Imaging Group, University of Twente, Enschede, 7500 AE, Netherlands.}}

\markboth{IEEE Transactions on Medical Imaging}%
{Boink \MakeLowercase{\textit{et al.}}: Joint Photoacoustic Reconstruction and Segmentation with a Partially Learned Algorithm}

\maketitle

\begin{abstract}
In an inhomogeneously illuminated photoacoustic image, important information like vascular geometry is not readily available when only the initial pressure is reconstructed. To obtain the desired information, algorithms for image segmentation are often applied as a post-processing step. In this work, we propose to jointly acquire the photoacoustic reconstruction and segmentation, by modifying a recently developed partially learned algorithm based on a convolutional neural network. We investigate the stability of the algorithm against changes in initial pressures and photoacoustic system settings. These insights are used to develop an algorithm that is robust to input and system settings. Our approach can easily be applied to other imaging modalities and can be modified to perform other high-level tasks different from segmentation. The method is validated on challenging synthetic and experimental photoacoustic tomography data in limited angle and limited view scenarios. It is computationally less expensive than classical iterative methods and enables higher quality reconstructions and segmentations than state-of-the-art learned and non-learned methods.
\end{abstract}


\IEEEpeerreviewmaketitle
\vspace{-3.5mm}
\section{Introduction}
\begin{figure}[!ht]
\resizebox{\linewidth}{!}{%
\begin{tikzpicture}
\draw[dashed,draw=black!40,line width = 1] (-0.1,1.5) rectangle (9.3, -0.65);
\draw [->,draw=black!80,line width=1.5](0,0) -- (0.95,0);
\node[label=above:{\small data}] at (0.4,-0.15) {};
\draw [fill=white,draw=black!80,line width=1.0,text=black!80] (1,-0.5) rectangle (3,0.5) node[pos=.5] {\large \textit{FBP}};
\draw [->,draw=black!80,line width=1.5](2,1.4) -- (2,0.55);
\node[label={[align=left]right:{\small physics}},text=black!80] at (1.85,1.2) {};
\node[label={[align=left]right:{\small modelling}},text=black!80] at (1.85,0.9) {};
\draw [->,draw=black!80,line width=1.5](3,0) -- (5.05,0);
\node[label=above:\small reconstruction] at (4,-0.15) {};
\draw [fill=white,draw=black!80,line width=1.0] (5.1,-0.5) rectangle (7.1,0.5) node[pos=.5] {\large \textit{U-Net}};
\draw [->,draw=black!80,line width=1.5](7.1,0) -- (9.2,0);
\node[label=above:\small segmentation] at (8.05,-0.2) {};
\node[text=black,align=left] at (8.36,1.15) {\textbf{\textit {two-step}}};
\node[text=black,align=left] at (8.4,0.85) {\textbf{\textit {approach}}};
\draw[dashed,draw=black!40,line width = 1] (-0.1,-0.85) rectangle (9.3, -3.0);
\draw [->,draw=black!80,line width=1.5](0,-1.5) -- (0.95,-1.5);
\node[label=above:{\small data}] at (0.4,-1.65) {};
\draw [fill=white,draw=black!90,line width=1.0,align=center,text=black!80] (1,-2) rectangle (7.1,-1) node[pos=.5] {\large \textit{Learned} \\ \large \textit{Primal-Dual}};
\draw [->,draw=black!80,line width=1.5](2,-2.9) -- (2,-2.05);
\node[label={[align=left]right:{\small physics}},text=black!80] at (1.85,-2.45) {};
\node[label={[align=left]right:{\small modelling}},text=black!80] at (1.85,-2.75) {};
\draw [->,draw=black!80,line width=1.5](7.1,-1.25) -- (9.2,-1.25);
\node[label=above:\small reconstruction] at (8.1,-1.40) {};
\draw [->,draw=black!80,line width=1.5](7.1,-1.75) -- (9.2,-1.75);
\node[label=above:\small segmentation] at (8.05,-1.95) {};
\node[text=black,align=left] at (8.0,-2.4) {\textbf{\textit {our}}};
\node[text=black,align=left] at (8.4,-2.7) {\textbf{\textit {approach}}};
\end{tikzpicture}}
\caption{Example of a two-step approach, where a segmentation is acquired by first applying FBP, after which U-Net is used as a second step. In our approach, we incorporate the physics modelling in the joint learned algorithm, providing both reconstruction and segmentation.}
\label{fig:UNet_vs_LPD}
\vspace{-4mm}
\end{figure}
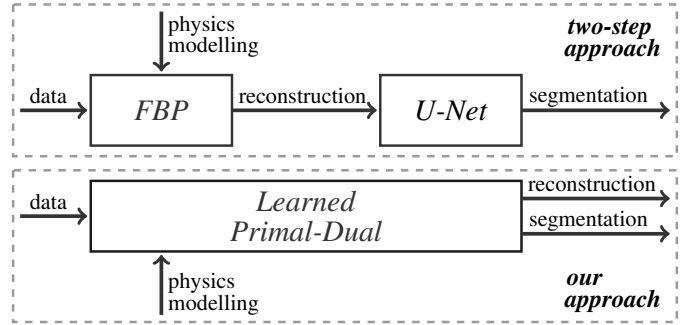

Photoacoustic tomography (PAT) is a hybrid imaging technique that combines high optical absorption contrast of tissues with high resolution from ultrasound detection. It is in the focus of research towards applications in various fields of biomedicine \cite{Zhou2016}, in particular for breast cancer imaging in humans \cite{Heijblom2015, Toi2017, Lin2018}. One of the main objectives in PAT is to acquire information on the vascular geometry in soft tissue: many diseases, such as breast cancer and rheumatoid arthritis can be characterised with increased blood vessel density and irregular vessel structure. A common practice is to first reconstruct the initial pressure image, after which a segmentation algorithm is applied to obtain the segmented vascular geometry. A big disadvantage of such `two-step' approaches is that reconstruction errors due to inaccurate physics modelling, noise or a lack of data will propagate to the subsequent segmentation step. In this work, we employ a partially learned algorithm to solve the reconstruction and segmentation task jointly. By utilising the {same} physics model for both problems, we can achieve higher quality segmentation results than classical non-learned or even learned two-step approaches (Fig. \ref{fig:UNet_vs_LPD}). {The algorithm is a partially learned one in the sense that the physics model is given, but other parts of the algorithm are learned}.

Common reconstruction methods for PAT are filtered backprojection (FBP) \cite{Kruger1995, Finch2004, Willemink2010, Haltmeier2014} and time reversal \cite{Burgholzer2007, Treeby2010}. These methods apply a variant of the adjoint acoustic operator on the measured data, which has been proven to give the exact result in case of infinite measurements without noise. Empirically, these approaches are still effective if `enough' measurements with limited noise are available. If this is not the case, these methods fail to provide adequate results and iterative approaches utilising regularisation \cite{Dean-Ben2012a, Huang2013, Arridge2016a, Arridge2016b, Boink2018, Nguyen2018, Frikel2018} are needed. It is not always clear which regulariser should be selected \cite{Boink2018}, which limits the practical use of those approaches. Moreover, iterative approaches often need many iterations to converge, which can make them computationally expensive.

Recently there has been a big interest in using deep neural networks for various tomography modalities to enable faster and higher quality reconstructions. Very deep networks like U-Net \cite{Ronneberger2015} are applied to post-process FBP-results and achieve a better reconstruction \cite{Jin2017, Antholzer2017, Kofler2018, Waibel2018}, but can also be used to obtain a segmentation of the tissue under consideration. These networks are not very effective when the quality of the input data is low, as is the case for under-sampled FBP reconstructions. For this reason, other algorithms incorporate the operator that describes the physics in the learning process. DALnet \cite{Schwab2018} is a non-iterative method that learns the weights of an operator that acts as a pseudo-inverse on the measured data. In \cite{Kelly2017}, the authors first solve a non-learned reconstruction problem, followed by a learned smoothing step; these steps are iteratively executed in the resulting algorithm. Many other learned iterative methods are based on classical iterative methods, but execute far fewer steps in their reconstruction procedure: variational networks \cite{Kobler2017b, Hammernik2017f, Hammernik2018} can be seen as a learned variant of proximal gradient methods, where kernel-function pairs of the regularisation term are learned. Learning the regularisation term is also the idea behind \cite{Chen2018} and \cite{Li2018}. In \cite{Adler2017a}, a learned variant of gradient-descent is applied, while learned proximal operators are investigated in \cite{Meinhardt2017}. The learned primal-dual (L-PD) algorithm \cite{Adler2018} can be seen as the learned variant of many popular and effective primal-dual methods \cite{Esser2010}. The key idea in \cite{Adler2017a}, \cite{Meinhardt2017} and \cite{Adler2018} is that instead of taking a pre-defined gradient or proximal step with respect to the regularisation function, an algorithm learns the steps that have to be taken to obtain a high-quality reconstruction. In \cite{Hauptmann2018}, L-PD was applied to 3D photoacoustic data. Since their 3D implementation of the acoustic operator is computationally expensive, the authors chose to not learn the whole `unrolled' algorithm, but learn per iteration instead. In case the acoustic operator is expensive, one can also choose to use an approximation of this operator \cite{Hauptmann2018b}, which still provides results superior to non-learned methods. Although empirically strong in their reconstruction capacity, most learned iterative methods lack a mathematical analysis of their convergence and stability. This makes them precarious for implementation in many applications, especially in clinical practice. A convergence proof for a method in which the regulariser was learned is given in \cite{Li2018}. Very recently, Banert \emph{et al.} \cite{Banert2018} presented a convergence proof for learned algorithms that make use of proximal operators.

Segmentation in biomedical imaging is a well-studied problem; see \cite{Acton2009} for an overview. New learned methods for segmentation are developed abundantly, of which U-Net \cite{Ronneberger2015} is currently a frequently used one. There are not many works on segmentation specifically designed for PAT; a simple segmentation procedure is given in \cite{Soetikno2012} and a complete `pipeline' from FBP reconstruction to segmentation is presented in \cite{Raumonen2018}. Recently, solving the reconstruction and segmentation problem jointly has become increasingly popular: by treating both problems simultaneously, reconstruction can benefit from information available in the segmentation and vice versa. One such method is given in \cite{Corona2018}, which also contains an overview on other non-learned joint methods. A recent work in which the segmentation is jointly solved with motion-estimation in an unsupervised learning approach is given in \cite{Qin2018}. \hl{A mathematical framework for solving high-level tasks together with reconstruction in a learned manner is provided in \cite{Adler2018b}. In that paper, the network architectures for reconstruction is not necessarily the same as that for the high-level task. }

In this paper, we make use of the partially learned L-PD method and modify it for the goal of joint reconstruction and segmentation in {two-dimensional} PAT. With this method, accurate information on soft tissue vasculature is directly available, without the necessity for post-processing. By using the acoustic model within the learning of the algorithm, both reconstruction and segmentation are of higher quality than in other learned algorithms. To obtain a robust method, the sensitivity of L-PD to image changes and changes in PAT system settings is investigated and solutions are provided. The method is tested on challenging synthetic and experimental data sets. 

To the best of our knowledge, this is the first time the joint problem of photoacoustic reconstruction and segmentation is addressed. Since a major interest in photoacoustic imaging is the visualisation of blood vessels, a segmentation of the vascular geometry is of high importance. Our method provides this segmentation, including a reliability estimate, which can be interpreted in an easy way. An analysis of the sensitivity of the learned iterative algorithm is performed, which has not been done before.

The remainder of this paper is organised as follows: in section \ref{sec:theory} we give a background to photoacoustic tomography, explain our forward model, and introduce the learned primal-dual method. In section \ref{sec:experiments}, we first present the neural network architecture that is used, before explaining how its sensitivity is investigated. In the same section, the joint segmentation-reconstruction method is provided, as well as the explanation of the experimental data on which it is tested. Quantitative and visual results of these experiments are given in section \ref{sec:results}. In section \ref{sec:conclusion} we conclude with some remarks and outlook for the future.
\section{Theory}\label{sec:theory}

\subsection{Photoacoustic tomography}
Photoacoustic signals are generated by illuminating tissue with nanosecond laser pulses, which causes the tissue to heat up at locations of optical absorption. Thermoelastic expansion causes the generation of pressure, which propagates in the form of ultrasound waves. An array of ultrasound detectors is placed around the tissue to detect this signal. For chosen wavelengths, certain tissue constituents such as haemoglobin have a higher optical absorption coefficient than surrounding tissue, giving the desired high contrast. The generated initial pressure $p_0$ depends linearly on the combination of optical absorption coefficient $\mu_a$ and fluence rate $\Phi$:
\begin{equation}
p_0\propto \mu_a\Phi.
\end{equation}

Quantitative PAT \cite{Cox2012} has the goal to reconstruct $\mu_a$, but has the drawback of being computationally expensive or inaccurate, depending on the light propagation model. The objective of the acoustic PAT inverse problem is to reconstruct $p_0$. This is computationally less expensive than the quantitative problem and the result already gives insight in the vascular geometry. However, a decaying fluence rate can cause problems for analysing this qualitative image, since the same blood vessel structure can have a lower intensity deeper inside the tissue. Therefore, there is a need for a computationally inexpensive method which does not only provide the reconstructed initial pressure, but also gives a segmentation of the vascular geometry, insensitive to fluence rate reduction in depth. 

\subsection{Forward model}\label{sec:fw_model}
In this section we mathematically describe the {two-dimensional} acoustic PAT forward problem. {Throughout the paper, with the exception of Appendix \ref{app:SoS}, we assume the sound speed to be constant and known. For the derivation of the model with non-constant sound speed, we refer to \cite[section 5.3]{Willemink2010}.} From this point onwards we write for the initial pressure $u$ instead of $p_0$ to improve readability of formulas and algorithms. We make use of a projection model with calibration as described in \cite{Wang2004}:
{\begin{equation}\label{eq:FW_full}
p(\mathbf{x},t) = |\mathbf{x}-\mathbf{x}_p|\left(\frac{1}{t}\int_{|\mathbf{x}-\tilde{\mathbf{x}}|=c(x)t} u(\tilde{\mathbf{x}})\text{d} \tilde{\mathbf{x}}\right)\ast_t p_\text{cal}\left(\mathbf{x},t'\right).
\end{equation}}
Here $p(\mathbf{x},t)$ is the measured pressure at location $\mathbf{x}$, which is the result of a spherical mean transform \cite{Kruger1995} applied to the initial pressure $u$ convolved with a calibration measurement $p_\text{cal}\left(\mathbf{x},t'\right)$. The last is the measurement of a point source located at $\mathbf{x}_p$, which is experimentally approximated. After pre-processing, we arrive at the simplified equation 
{\begin{equation}\label{eq:FW_prep}
f = Ku:=\int_{|\mathbf{x}-\tilde{\mathbf{x}}|=c(x)t} u(\tilde{\mathbf{x}})\text{d} \tilde{\mathbf{x}},
\end{equation}}
where $K$ is the acoustic forward operator mapping the initial pressure to the preprocessed measurements $f$. For a more detailed explanation of the calibration measurement and pre-processing, we refer to \cite[Chapter 2]{Willemink2010} and \cite{Boink2018}. \hl{For computational reasons, we make use of a matrix implementation of $K$. However, it should be noted that our algorithm can also be applied when a matrix implementation of the acoustic operator is not available.}

\subsection{From non-learned to learned primal-dual optimisation}\label{sec:LPD}
Many inverse problems, including the PAT reconstruction problem \eqref{eq:FW_prep}, can be written as 
\begin{equation}
f = Ku +\varepsilon,
\end{equation}
where data $f$ is known, an approximate forward model $K$ is available and the noise level of the additive noise $\varepsilon$ is estimated. In case a direct reconstruction method does not give an adequate result, variational methods are employed to solve the minimisation problem
\begin{equation}\label{eq:varmethod}
\min_u F_f(Au) + G(u).
\end{equation} 
Here $F_f$ is an operator that acts on the dual domain and handles the data, and $G$ is an operator that acts on the primal domain. For instance, writing down a classical $L^2-TV$ model \cite{Rudin1992} in discrete form, it reads
\begin{equation}\label{eq:TV}
\min_u \frac12\norm{Ku-f}_2^2 + \alpha\norm{\nabla u}_1.
\end{equation} 
This can be put in the form of \eqref{eq:varmethod} by choosing $A:=[K,\nabla]$ and for $F_f$ choose a combination of the 2-norm and 1-norm acting on $Au$, while $G(u):=0$. The reason of writing it down in such a form is that any primal-dual algorithm can be applied to solve this minimisation problem \cite{Esser2010}, of which primal-dual hybrid gradient (PDHG) \cite{Chambolle2011} is a popular choice for tomography problems \cite{Sidky2012, Boink2018}. A key element in primal-dual algorithms is updating via the proximal operator, which for a general scaled functional $\gamma H(u)$ is defined as
\begin{equation}
\prox_{\gamma H(u)} = \argmin_{\tilde{u}}\left\{\frac12\norm{u-\tilde{u}}_2^2 + \gamma H(\tilde{u})\right\}.
\end{equation}
This proximal operator takes a descent step in the functional $H(\tilde{u})$, while staying close to a previous iterate $u$. 

\vspace{3mm}
\fbox{\begin{minipage}{0.95\linewidth}
\begin{center}
\begin{algorithmic}
\For{$n \gets 1$ to $\tilde{N}$}
\State $q^{n+1} = \prox_{\sigma F^*_f}\Big(q^n+\sigma A\left[(1+\theta)u^{n}-\theta u^{n-1}\right]\Big),$
\State $u^{n+1} = \prox_{\tau G} \Big(u^{n}-\tau A^* q^{n+1}\Big).$
\EndFor
\end{algorithmic}
\end{center}
\end{minipage}}
\vspace{1mm}
\begin{minipage}{\linewidth}
$~$\vspace{-1mm}\captionof{algorithm}{PDHG (non-learned).\label{alg:PDHG} $~$\vspace{4mm}}
\end{minipage}

In Algorithm \ref{alg:PDHG}, PDHG is stated, where the $\theta$-acceleration step is applied in the first update. The first update takes a proximal step with respect to the convex conjugate of $F$ and acts in the dual domain; its inputs are the previous dual update $q^n$, data $f$ and the acoustic operator $A$ applied to the primal updates $u^{n-1}$ and $u^n$. The second update takes a proximal step with respect to $G$ and acts on the previous primal update $u^n$ and the adjoint operator $A^*$ applied to the dual update $q^{n+1}$. {The adjoint operator $A^*$ is obtained by first discretising $A$ in a matrix and then taking the numerical adjoint.}

\begin{figure*}
\resizebox{1.025\linewidth}{!}{%
\begin{tikzpicture}
\begin{scope}[xshift=23cm,yshift=2.5cm,scale=0.9, every node/.style={transform shape}]
\input{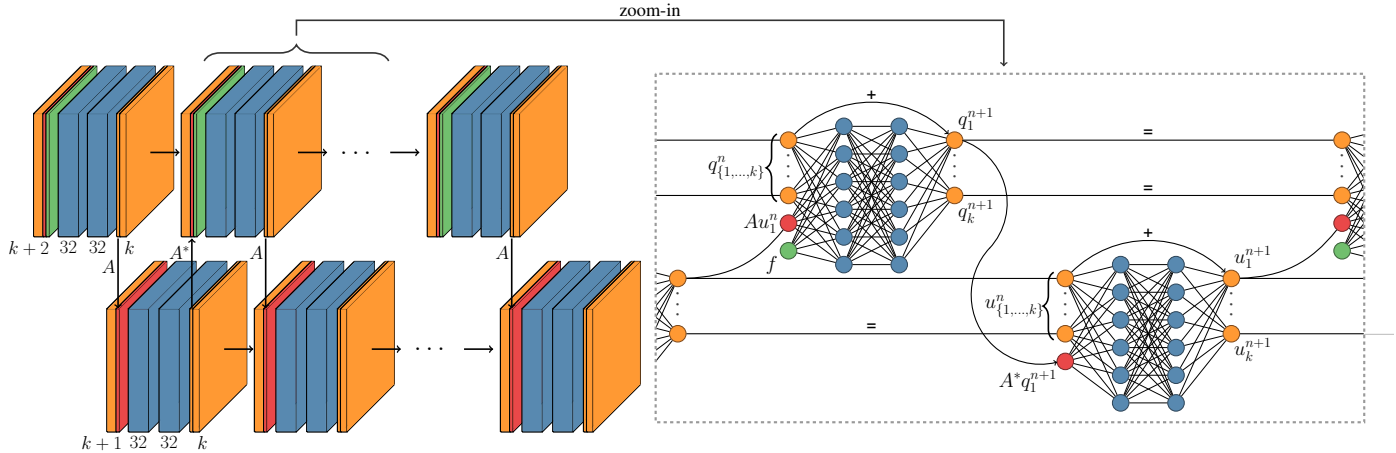}
\end{scope}

\definecolor{bluee}{RGB}{88,137,176}
\definecolor{greenn}{RGB}{113,191,110}
\definecolor{orangee}{RGB}{255,153,51}
\definecolor{redd}{RGB}{233,72,73}
\definecolor{darkbluee}{RGB}{44,68,88}
\definecolor{darkgreenn}{RGB}{56,95,55}
\definecolor{darkorangee}{RGB}{126,72,25}
\definecolor{darkredd}{RGB}{116,36,36}

\newcommand{\networkLayer}[6]{
	\def\a{#1} 
	\def\b{0.03}
	\def\c{#2} 
	\def\t{#3} 
	\def\d{#4} 

	\draw[line width=0.5mm](\c+\t,0,\d) -- (\c+\t,\a,\d) -- (\t,\a,\d);                                                      
	\draw[line width=0.5mm](\t,0,\a+\d) -- (\c+\t,0,\a+\d) node[midway,below] {#6} -- (\c+\t,\a,\a+\d) -- (\t,\a,\a+\d) -- (\t,0,\a+\d); 
	\draw[line width=0.5mm](\c+\t,0,\d) -- (\c+\t,0,\a+\d);
	\draw[line width=0.5mm](\c+\t,\a,\d) -- (\c+\t,\a,\a+\d);
	\draw[line width=0.5mm](\t,\a,\d) -- (\t,\a,\a+\d);

	\filldraw[#5] (\t+\b,\b,\a+\d) -- (\c+\t-\b,\b,\a+\d) -- (\c+\t-\b,\a-\b,\a+\d) -- (\t+\b,\a-\b,\a+\d) -- (\t+\b,\b,\a+\d); 
	\filldraw[#5] (\t+\b,\a,\a-\b+\d) -- (\c+\t-\b,\a,\a-\b+\d) -- (\c+\t-\b,\a,\b+\d) -- (\t+\b,\a,\b+\d);

	\ifthenelse {\equal{#5} {}}
	{} 
	{\filldraw[#5] (\c+\t,\b,\a-\b+\d) -- (\c+\t,\b,\b+\d) -- (\c+\t,\a-\b,\b+\d) -- (\c+\t,\a-\b,\a-\b+\d);} 
}
		

\newcommand{\networkLayerS}[6]{
	\networkLayer{#1}{#2}{1.455*#4+#3}{2.75*#4}{#5}{#6}
}

\networkLayerS{4.0}{0.3}{0.0}{0.0}{color=orangee}{\LARGE{$k+2$ $~~$}}  
\networkLayerS{4.0}{0.1}{0.3}{0.0}{color=redd}{}        
\networkLayerS{4.0}{0.1}{0.4}{0.0}{color=greenn}{}    

\networkLayerS{4.0}{0.65}{0.8}{0.0}{color=bluee}{\LARGE{$32$}}  

\networkLayerS{4.0}{0.65}{1.75}{0.0}{color=bluee}{\LARGE{$32$}}  
\networkLayerS{4.0}{0.1}{2.7}{0.0}{color=orangee}{}  
\networkLayerS{4.0}{0.2}{2.8}{0.0}{color=orangee}{\LARGE{$~$ $k$}}

\networkLayerS{4.0}{0.3}{0.0}{6.0}{color=orangee}{\LARGE{$k+1$ $~~$}}  
\networkLayerS{4.0}{0.1}{0.3}{6.0}{color=redd}{}        

\networkLayerS{4.0}{0.65}{0.7}{6.0}{color=bluee}{\LARGE{$32$}}  

\networkLayerS{4.0}{0.65}{1.7}{6.0}{color=bluee}{\LARGE{$32$}}  

\networkLayerS{4.0}{0.1}{2.7}{6.0}{color=orangee}{}  
\networkLayerS{4.0}{0.2}{2.8}{6.0}{color=orangee}{\LARGE{$~$ $k$}}  

\draw [line width=1.5pt,->] (1.2,-1.6) -- (1.2,-3.88) {};
\node[text=black,align=left] at (0.9,-2.5) {\LARGE \textit {$A$}};
\draw [line width=1.5pt,->] (2.25,1.2) -- (3.2,1.2) {};

\networkLayerS{4.0}{0.3}{4.8}{0.0}{color=orangee}{}  
\networkLayerS{4.0}{0.1}{5.1}{0.0}{color=redd}{}        
\networkLayerS{4.0}{0.1}{5.2}{0.0}{color=greenn}{}    

\networkLayerS{4.0}{0.65}{5.6}{0.0}{color=bluee}{}  

\networkLayerS{4.0}{0.65}{6.55}{0.0}{color=bluee}{}  
\networkLayerS{4.0}{0.1}{7.5}{0.0}{color=orangee}{}  
\networkLayerS{4.0}{0.2}{7.6}{0.0}{color=orangee}{}

\draw [line width=1.5pt,->] (3.6,-3.88) -- (3.6,-1.6) {};
\node[text=black,align=left] at (3.2,-2.05) {\LARGE \textit {$A^*$}};
\draw [line width=1.5pt,->] (4.65,-5.2) -- (5.6,-5.2) {};

\networkLayerS{4.0}{0.3}{4.8}{6.0}{color=orangee}{}  
\networkLayerS{4.0}{0.1}{5.1}{6.0}{color=redd}{}        

\networkLayerS{4.0}{0.65}{5.5}{6.0}{color=bluee}{}  

\networkLayerS{4.0}{0.65}{6.5}{6.0}{color=bluee}{}  

\networkLayerS{4.0}{0.1}{7.5}{6.0}{color=orangee}{}  
\networkLayerS{4.0}{0.2}{7.6}{6.0}{color=orangee}{}  

\draw [line width=1.5pt,->] (6.0,-1.6) -- (6.0,-3.88) {};
\node[text=black,align=left] at (5.7,-2.05) {\LARGE \textit {$A$}};
\node[text=black,align=left] at (5.8,-2.15) {};

\draw [line width=1.5pt,->] (7.05,1.2) -- (8.0,1.2) {};
\node[text=black,align=left] at (9.0,1.2) {\Huge $\dots$};
\draw [line width=1.5pt,->] (10.05,1.2) -- (11.0,1.2) {};

\draw [line width=1.5pt,->] (9.45,-5.2) -- (10.4,-5.2) {};
\node[text=black,align=left] at (11.4,-5.2) {\Huge $\dots$};
\draw [line width=1.5pt,->] (12.45,-5.2) -- (13.4,-5.2) {};

\networkLayerS{4.0}{0.3}{12.8}{0.0}{color=orangee}{}  
\networkLayerS{4.0}{0.1}{13.1}{0.0}{color=redd}{}        
\networkLayerS{4.0}{0.1}{13.2}{0.0}{color=greenn}{}    

\networkLayerS{4.0}{0.65}{13.6}{0.0}{color=bluee}{}  

\networkLayerS{4.0}{0.65}{14.55}{0.0}{color=bluee}{}  
\networkLayerS{4.0}{0.1}{15.5}{0.0}{color=orangee}{}  
\networkLayerS{4.0}{0.2}{15.6}{0.0}{color=orangee}{}

\networkLayerS{4.0}{0.3}{12.8}{6.0}{color=orangee}{}  
\networkLayerS{4.0}{0.1}{13.1}{6.0}{color=redd}{}        

\networkLayerS{4.0}{0.65}{13.5}{6.0}{color=bluee}{}  

\networkLayerS{4.0}{0.65}{14.5}{6.0}{color=bluee}{}  

\networkLayerS{4.0}{0.1}{15.5}{6.0}{color=orangee}{}  
\networkLayerS{4.0}{0.2}{15.6}{6.0}{color=orangee}{}

\draw [line width=1.5pt,->] (14.0,-1.6) -- (14.0,-3.88) {};
\node[text=black,align=left] at (13.7,-2.05) {\LARGE \textit {$A$}};

\draw[decoration={brace,raise=5pt,amplitude=15pt},draw=black!80,decorate,-,line width=1.5pt]
  	(4.0,4.0) -- (10.0,4.0) {};

\draw [line width=1.5pt,->,draw=black!80] (7.0,5) to (7.0,5.5) to node [auto] {\LARGE zoom-in} (30,5.5) to (30,4.0);
\end{tikzpicture}}
\caption{Visualisation of learned primal-dual architecture as described in Algorithm \ref{alg:LPD}. Number of channels is indicated below the layers on the left. On the right, every node corresponds to a $3\times3$ convolution followed by a ReLU-activation function.}
\label{fig:recon_segm_setup}
\vspace{-2.5mm}
\end{figure*}

\vspace{3mm}
\fbox{\begin{minipage}{0.95\linewidth}
\begin{center}
\begin{algorithmic}
\For{$n \gets 1$ to $N$}
\State $q^{n+1}_{\{1,\dotsc,k\}} = q^n_{\{1,\dotsc,k\}} + \Gamma_{\Theta_n}\left(q^n_{\{1,\dotsc,k\}}, A u^n_1,f\right),$
\State $u^{n+1}_{\{1,\dotsc,k\}} = u^{n}_{\{1,\dotsc,k\}} + \Lambda_{\Psi_n} \left(u^n_{\{1,\dotsc,k\}}, A^* q^{n+1}_1\right).$
\EndFor
\end{algorithmic}
\end{center}
\end{minipage}}\\
\begin{minipage}{\linewidth}
$~$\vspace{-1mm} {\captionof{algorithm}{L-PD (learned). See Fig. \ref{fig:recon_segm_setup} for a visualisation. \label{alg:LPD}\vspace{4mm}}}
\end{minipage}

In this work, we make use of the learned primal-dual (L-PD) algorithm \cite{Adler2018}, which can be derived from PDHG. The idea for the L-PD approach (Algorithm \ref{alg:LPD}) is to not choose the functionals $F$ and $G$ explicitly, but learn the best update steps for each iteration. This is achieved by a CNN, here represented by the nonlinear functions {$\Gamma_{\Theta_n}$ and $\Lambda_{\Psi_n}$, where $\Theta_n$ and $\Psi_n$} describe the learned weights for iteration $n$. The key idea is that the inputs of the learned steps of Algorithm \ref{alg:LPD} are still the same as in Algorithm \ref{alg:PDHG}, such that the primal-dual structure is preserved. Note that weights of the network can be different for every iteration $n\in\{1,\dots,N\}$. Moreover, instead of updating one channel of the primal and dual, we allow the network to use $k$ channels. This could encode some kind of `history', which is similar to the acceleration of PDHG, where not only $u^n$, but also $u^{n-1}$ is an input in the first proximal step. However, the different channels can represent more general structures in the primal and dual domain that are useful for reconstruction. {Channel 1 of the final iteration $N$ gives our desired reconstruction. This leaves room for the other channels to represent other useful information; this will be leveraged in section \ref{sec:segmentation}, where we use channel 2 to provide a non-binary segmentation.} In section \ref{sec:results} it will be shown that the L-PD algorithm provides higher quality reconstructions than PDHG applied to a TV regularised model \eqref{eq:TV}. An additional advantage is that L-PD can be enforced to only use a small amount of iterations, whereas in PDHG, one has to wait till convergence, which generally takes many more iterations. {A visualisation of the algorithm as a neural network is given in Fig. \ref{fig:recon_segm_setup}.}

\section{Methodology}\label{sec:experiments}
A major interest in photoacoustic tomography is the visualisation of blood vessels in tissue as a marker of diseases, for instance in breast imaging. For this reason, we made use of retinal blood vessel images from the openly available DRIVE-dataset \cite{Staal2004}. A total of 768 training images and 192 test images were obtained by preprocessing patches with a size of $192\times192$ pixels from this dataset. These training and test images have been further processed to contribute to our two goals: obtaining robustness to variations in images and system settings, and obtaining an additional segmentation in a realistic photoacoustic scenario. This is explained in detail in sections \ref{sec:robustness} and \ref{sec:segmentation} respectively. \hl{The discrete size of the synthetic measurements is $1450\times n_s$ pixels, where $n_s$ is the number of sensors used.}

For the L-PD algorithm, both a small network and a larger network were trained. In the first one, the number of iterations was chosen to be $N=10$ and the number of primal- and dual channels $k=5$ (cf. Algorithm \ref{alg:LPD}). In the second one, $N=5$ and $k=2$ were chosen. In both networks, there are 32 channels in 2 hidden layers. The filter size for the convolutions is $3\times3$ and ReLUs are chosen as activation functions. The Adam optimiser \cite{Kingma2015} with a mean-squared-error loss on the difference between ground truth and reconstruction is used: 
\begin{equation}\label{eq:Loss_1}
L=\min_{\Theta_{\{1,\dots,N\}}}\norm{u_{GT}-u_1^N}^2_2.
\end{equation}
For stability in the optimisation, the batch size increases from 2 to 16 in three steps \cite{Smith2017} during 200 Epochs. {We make use of TensorFlow, executed on a single GTX 1080 TI GPU. Training took between 1 and 4 hours, depending on the amount of detectors used. It should be noted that this depends heavily on the specific operator $A$ used: for training, one has to perform $2N$ iterations per backprojection, which makes the application of $A$ the limiting factor for most tomography problems. 

After training, the computational cost for applying the algorithm also depends mostly on the operator $A$. For each iteration, $A$ and $A^*$ have to be applied once. This means that the computational cost is approximately $2N$ times that of FBP. For a TV-reconstruction, the final iteration $\tilde{N}$ depends on the desired accuracy, but is usually in the order of 100 or 1000 (c.f. section \ref{sec:comparison}), while we choose $N=5$ or $N=10$ for L-PD.}

\subsection{Training for robustness}\label{sec:robustness}
In this experiment we investigate robustness of the L-PD network to changes in the image and system settings, and give solutions to improve this robustness. Digital phantoms are created by setting the background of the vascular images to 0 and setting the maximum to 1. Synthetic data is generated by applying the forward model as explained in section \ref{sec:fw_model} to these phantoms. For the standard data set, a setting of 32 detectors is used, uniformly placed around the phantom {(c.f. right of Fig. \ref{fig:exp_photo})}. The L-PD algorithm is trained with the architecture as specified above. 

\subsubsection{Image uncertainty}
\begin{figure}[!ht]
\centering
\includegraphics[width=\linewidth]{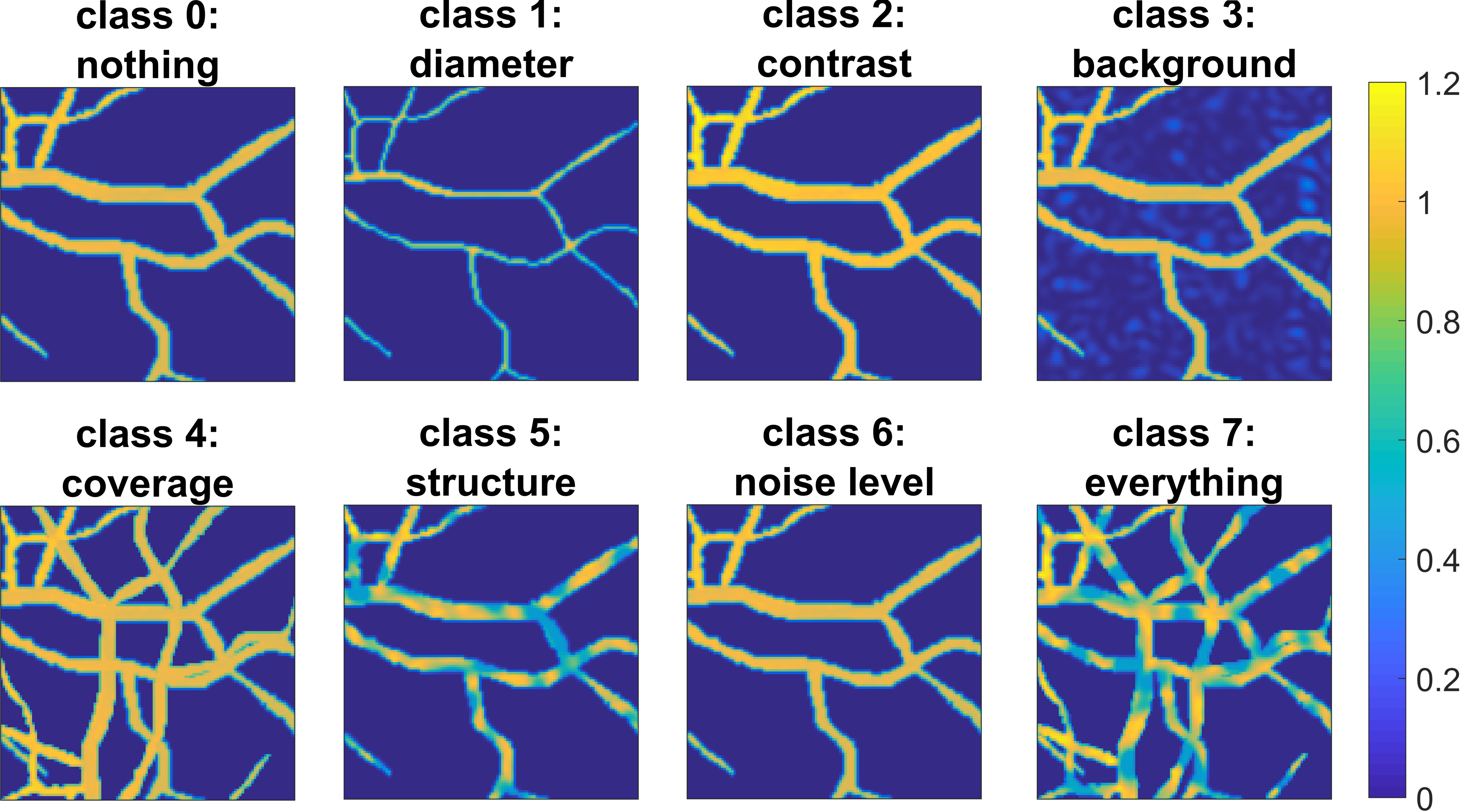}
\caption{Example image where one feature (class 1-6) or more features (class 7) are changed. See Table \ref{tab:image_changes} for detailed information.}
\label{fig:image_changes}
\end{figure}

In Table \ref{tab:image_changes}, 8 classes of images are defined, in which one or multiple image properties have been changed. For a visual impression, one instance of every class has been shown in Fig. \ref{fig:image_changes}. To analyse the sensitivity of L-PD, the algorithm was trained on class 0 and applied to the test set of class $i$, with $i=1,2,\dots,7$. Then the network was retrained on a training set of class $i$ and again applied to the test set. We investigate how this affects the reconstruction quality. 

\begin{table*}[ht!]
\vspace{2mm}
\caption {Explanation of image classes where one feature (class 1-6) or multiple features (class 7) are changed. See Fig. \ref{fig:image_changes} for a visual impression.}
\vspace{-3mm}
\begin{center}
{\rowcolors{2}{white}{gray!25}
\bgroup
\def\arraystretch{1.15}
\begin{tabular}{ l | l | l } 
 \hline
 \textbf{class ($c_i$)} & \textbf{feature change} & \textbf{explanation} \\ \hline \hline
 0 & nothing & \textit{For an example image see bottom left image in Fig. \ref{fig:image_changes}.}\\
 1 & diameter & Change of vessel diameter, where the pixel-wise increase is randomly chosen from the set $\{-2,-1,0,1,2\}$. \\ 
 2 & contrast & Vessel intensity randomly chosen from the set $\{0.5, 0.7, 1, 1.4, 2\}$. \\
 3 & background & With probability $p=\frac12$ a uniform increase in background intensity, with $p=\frac12$ a speckled background. \\ 
 4 & coverage & With probability $p=\frac13$ a removal of all the smallest vessels, with $p=\frac13$ nothing, with $p=\frac13$ a doubling of vessels.\\
 5 & structure & Inhomogeneous vessel intensity scaled between $[m,1]$, where $m$ is randomly chosen from $\{0.2, 0.4, 0.6, 0.8, 1\}$.  \\ 
 6 & noise level & Gaussian noise level randomly chosen from $[0,3\sigma]$, where $\sigma$ is the standard noise level \\
 7 & All of the above & all of the above changes have all been applied with given probabilities.\\ 
 \hline
\end{tabular}
\egroup
}
\end{center}\label{tab:image_changes}
\vspace{-3mm}
\end{table*} 

\subsubsection{System uncertainty}\label{sec:system_uncertainty}
To investigate the sensitivity of L-PD to changes in system settings, the algorithm is first trained on data from 32 uniformly placed detectors. The trained algorithm is applied to test data with 16 or 64 uniformly placed detectors and 32 detectors in a limited view scenario {(c.f. right of Fig. \ref{fig:exp_photo})}. Next, the network is retrained on a training set where the detectors were randomly chosen from a possibility of 128 detectors. We investigate how retraining on this more diverse training set affects the reconstruction quality. {It is possible to explicitly treat the limited view problem by weighing the pixels at different locations differently \cite{Paltauf2007,Schwab2018}. Instead, we choose to let the L-PD algorithm learn to correct for the limited view problem.

In Appendix \ref{app:SoS} and Appendix \ref{app:cal} robustness to heterogeneous sound speed changes and uncertainty in the calibration measurements are analysed, respectively.}

\subsection{Joint reconstruction and segmentation}\label{sec:segmentation}
The experiments in section \ref{sec:robustness} are meant to analyse the robustness of the L-PD algorithm to changes in image or in system settings. In these experiments, a homogeneous internal fluence rate is assumed, so that the initial pressure within the blood vessels is uniform. 

This is not a realistic scenario, especially for breast imaging, since the fluence rate drops rapidly with depth from the illuminated surface due to scattering and absorption in tissue. To get more realistic digital phantoms, we model light propagation with the diffusion approximation (DA) of the radiative transfer equation (RTE) \cite{Schweiger1993}. This model is applied to the digital phantoms of class 0 in \ref{sec:robustness}, which serve as the optical absorption coefficient $\mu_a$. The DA has been implemented in the FEniCS framework \cite{Alnaes2014} in Python. In the test set, 48 images have homogeneous internal fluence rate and 144 images have a heterogeneous internal fluence rate modelled with the DA with illumination from either 1, 2 or 4 sides. 

For the test set, values for absorption $\mu_a$ and reduced scattering $\mu_s'$ were selected from literature \cite{Jacques2013} to represent the values of whole blood for the vessels and glandular tissue or lipid for the background. For the training set, vessel absorption and background scattering were chosen in a wider range to improve reconstruction robustness, as can be seen in section \ref{sec:results_robustness}. The exact values are stated in Table \ref{tab:opt_prop}.

\begin{table}[ht!]
\vspace{2mm}
\caption {Optical properties in the test set.}
\vspace{-3mm}
\begin{center}
\bgroup
\def\arraystretch{1.15}
\begin{tabular}{ l | l | l } 
 \hline
\textbf{data set} & \textbf{part of phantom} & \textbf{value}\\
\hline \hline
\cellcolor{gray!10}& \cellcolor{gray!25} & \cellcolor{gray!25}$\mu_a=0.40$ mm$^{-1}$\\ 
\cellcolor{gray!10}& \multirow{-2}{*}{\cellcolor{gray!25} vessel} & \cellcolor{gray!25}$\mu_s'=0.45$ mm$^{-1}$\\ \cline{2-3}
\cellcolor{gray!10}& \multirow{2}{*}{background} & $\mu_a=0.004$ mm$^{-1}$\\ 
\multirow{-4}{*}{\cellcolor{gray!10} Test set}& & $\mu_s'=0.97$ mm$^{-1}$\\
 \hline \hline
\cellcolor{gray!10} & \cellcolor{gray!25} & \cellcolor{gray!25}$\mu_a\in\big[0.20$ mm$^{-1}, 0.60$ mm$^{-1}\big]$\\ 
\cellcolor{gray!10} & \multirow{-2}{*}{\cellcolor{gray!25} vessel} & \cellcolor{gray!25}$\mu_s'=0.45$ mm$^{-1}$\\ \cline{2-3}
\cellcolor{gray!10}& \multirow{2}{*}{background} & $\mu_a=0.004$ mm$^{-1}$\\ 
\multirow{-4}{*}{\cellcolor{gray!10} Training set} & & $\mu_s'\in\big[0.50$ mm$^{-1}, 2.00$ mm$^{-1}\big]$\\
 \hline
\end{tabular}
\egroup
\end{center}\label{tab:opt_prop}
\vspace{-4mm}
\end{table} 

For these more realistic phantoms with spatially varying fluence rate it is difficult to detect the vascular geometry, since blood vessels deep in the tissue will give a lower initial pressure than shallow ones. Moreover, low intensity parts of the reconstruction can be invisible due to artefacts originating from higher intensity parts. To accurately reveal the vascular geometry at all tissue depths, we propose a method to compute the segmentation jointly with the photoacoustic reconstruction: instead of solely training for the best reconstruction, the neural network architecture is also trained for a segmentation output. {The same network as in section \ref{sec:robustness} is used, which was explained in section \ref{sec:LPD}. Now however, not only a reconstruction output $u_1^N$ in the first channel is required, but also a segmentation output $u_2^N$ in the second channel of iteration $N$.} This is achieved by extending the reconstruction-based loss function \eqref{eq:Loss_1} with a binary cross-entropy loss on a segmentation output compared to the ground truth segmentation $u_S$.
\begin{align}\label{eq:Loss_2}
L=\min_{\Theta_{\{1,\dots,N\}}}\Big\{&\norm{u_{GT}-u_1^N}^2_2 \\- \beta &\Big(u_S\log(u_2^N)+(1-u_S)\log(1-u_2^N)\Big)\Big\}.\nonumber
\end{align}
Here $\beta$ is a parameter that determines the weighting between reconstruction quality and segmentation quality. For all our experiments, $\beta=0.5$ was chosen, as empirically it is seen that solutions are not very sensitive to this parameter. {Note that all the layers before the output layer are the same, meaning that the same information is used for the reconstruction and for the segmentation, which makes it a joint algorithm.}

Because the neural network needs to be differentiable, the segmentation output $u_2^N$ is not binary, but in the range $[0,1]$. This can be interpreted as some kind of reliability estimate: when a pixel value is very close to 1, it is almost surely part of a vessel, while it is almost surely not when its value is very close to 0. To obtain a truly binary segmentation, the non-binary output will be thresholded at the value that gives the highest segmentation accuracy and highest Dice-coefficient \cite{Zijdenbos1994}. The interpretation of the output as a reliability estimate implies that this threshold should be similar for all outputs, which is verified by experiments. For this reason, we choose one threshold which provides the best binary segmentation, averaged over the whole training set.

\subsection{Comparison with other methods}\label{sec:comparison}
{The reconstruction output of the joint L-PD algorithm is compared with the following other methods:
\begin{itemize}
\item Direct FBP method as described in \cite{Willemink2010}.
\item Iterative TV-regularised method \eqref{eq:TV} using PDHG (Algorithm \ref{alg:PDHG}) as described in \cite{Boink2018}. To ensure that the algorithm is converged to high precision, we use 1000 iterations. Visually, the results do not change after the first 100 iterations. The regularisation parameter $\alpha$ was chosen such that the PSNR between ground truth and reconstruction is highest. Moreover, a scaling was introduced to compensate for contrast loss that is well known for TV. \hl{That is, the reconstruction is multiplied with a scalar such that the outcome gives the highest PSNR.}
\item Learned U-Net approach \cite{Jin2017}, where a neural network is used to post-process the FBP-reconstruction for a higher quality reconstruction. Parameters are chosen as in \cite{Jin2017}. 
\end{itemize}
For segmentation, the following comparisons are made:
\begin{itemize}
\item Thresholding the FBP-reconstruction and TV-reconstruction.
\item A spectral globally convex segmentation (S-GCS) \cite{Zeune2017}, which can be seen as a convexified version of the level set method \cite{Goldstein2010} applied with different scales. We use the numerical implementation as described in \cite{Zeune2017}, where we choose 15 scales of $\alpha\in\{2,3,\dots,16\}$.   The output scales are normalised to make comparison to the L-PD output possible. The algorithm is applied to the FBP-reconstruction and uses an $L^1$ data-fidelity term: this can potentially account for the decaying intensity in the reconstruction. 
\item Learned U-Net approach \cite{Ronneberger2015}, where we ask for a segmentation with the FBP-reconstruction as input.
\end{itemize}
}

\subsection{Application to experimental data}\label{sec:exp_phantom}
{We apply the L-PD algorithm that was trained on the training set as explained in section \ref{sec:segmentation} to experimental data. With this we check the potential of L-PD to be applied to real data and the uncertainty that comes with it.}

\subsubsection{Experimental setup}
\begin{figure}[!ht]
\centering
\includegraphics[width=\linewidth]{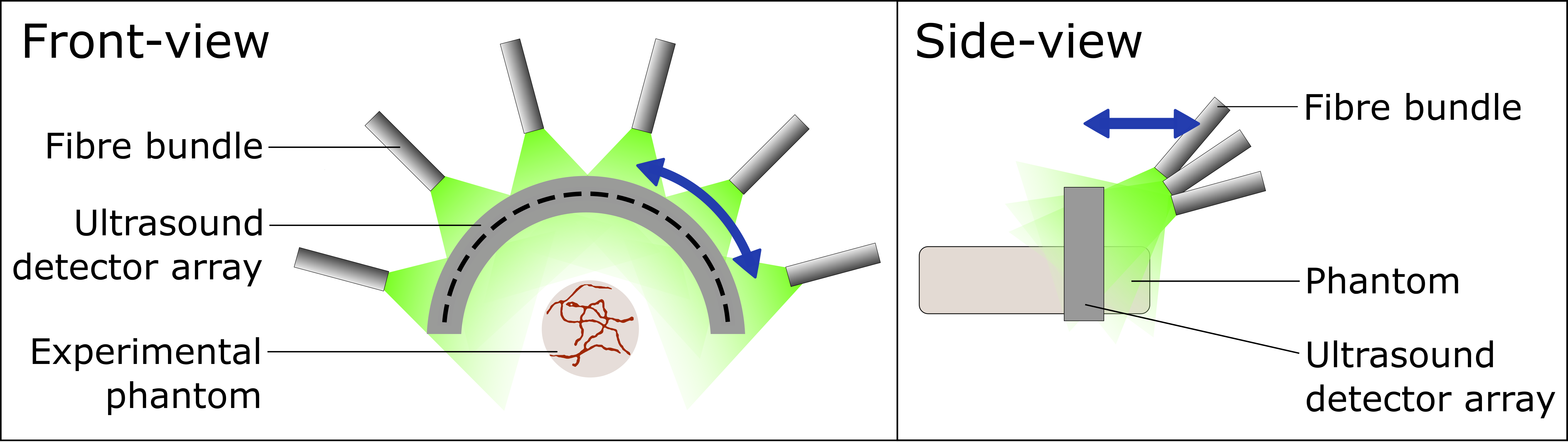}
\caption{Schematic overview of the experimental setup, where a phantom is measured.}
\label{fig:exp_setup}
\vspace{-5mm}
\end{figure}

We make use of a tomographic photoacoustic imager as specified in \cite{Es2015}. A laser delivers 5 ns pulses of optical energy with a wavelength of 532 nm. For the recording of pressure waves a 1D piezoelectric detector array with 64 elements in a half-circle is used. The detector array has a central frequency of 7.5 MHz with a fractional bandwidth of 85\%. \hl{The acquired data is discretised with a time sampling of 25 ns, for a total time of approximately 45 $\mu$s.} Both the detectors and fibre bundles are simultaneously rotatable over 360 degrees. They can also be translated in order to image multiple slices. The detector array has a narrow focus (0.6 mm slice thickness) in one dimension, making it suitable for 2D slice based imaging. A schematic overview of the experimental setup is shown in Figure \ref{fig:exp_setup}. For a more extensive explanation of the setup, we refer to \cite{Es2015}. 

\subsubsection{Vascular phantom}
\begin{figure}[ht!]
\resizebox{\linewidth}{!}{%
\begin{tikzpicture}
\input{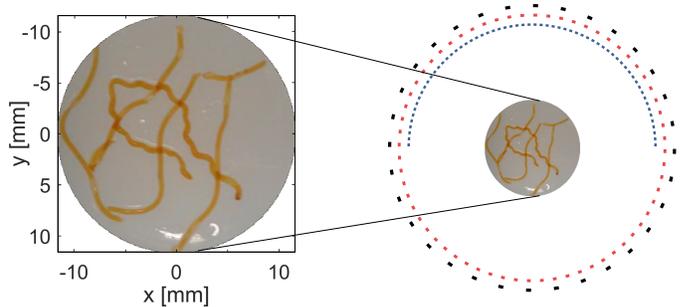}
\end{tikzpicture}}
\caption{\textbf{Left:} Experimental phantom consisting of absorbing filaments in a scattering medium. \textbf{Right:} Schematic of detector settings used in experiments. Black: 32 detectors uniform. Red: 64 detectors uniform. Blue: 64 detectors limited-view.}
\label{fig:exp_photo}
\end{figure}

For experimental data, a phantom with vessel-shaped absorbers in an optically scattering medium was created. Absorbing filaments were gently placed on stiff agar gel, after which a second layer of agar solution was poured on the whole and allowed to harden. A photograph of the filaments before pouring the second layer of agar is shown in Fig. \ref{fig:exp_photo}. {The experimental agar phantom has the property that the sound speed is almost identical to that of water of the same temperature.} For details on the phantom creation, we refer to \cite{Boink2018}. 

Data is obtained in a limited-view and a uniform detector setting: in the former, fibre bundles and detector array are placed on the top of the phantom; in the latter, fibre bundles and detector array are rotated over an angle of 180 degrees, which means that the detector array covers the whole circle. We point out that by this rotation not only acquisition is uniform, but also light is more homogeneously spread over the surface of the phantom. A schematic drawing of the detector setting is given in Fig. \ref{fig:exp_photo}, with two uniform settings (32 and 64 detectors) and one limited-view setting (64 detectors). 

\section{Results}\label{sec:results}
\begin{figure}[ht!]
\centering
\includegraphics[width=0.9\linewidth]{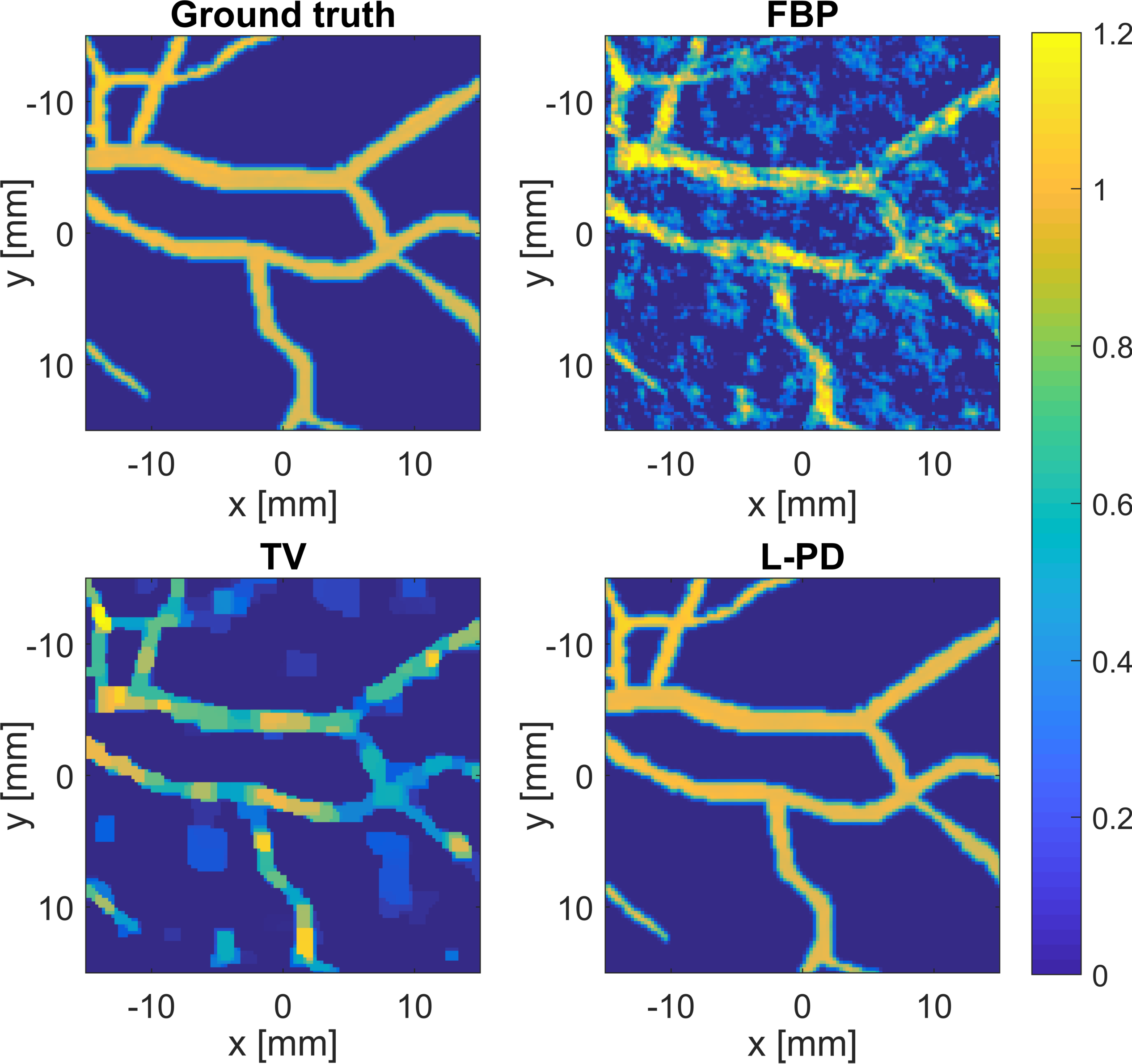}
\caption{Reconstructions of a vascular phantom of class 0 using a uniform placement of 32 detectors.}
\label{fig:plain_result}
\vspace{-3mm}
\end{figure}

In Fig. \ref{fig:plain_result} reconstructions of a digital phantom from class 0 have been shown, where data was obtained by simulating for a uniform placement of 32 detectors. In line with the results in \cite{Adler2018} it can be seen that the non-learned methods FBP and TV show artefacts in the background due to limited measurements and noise, while L-PD gives an almost perfect reconstruction with clear contrast and zero background.

\subsection{Robustness against uncertainty}\label{sec:results_robustness}
\subsubsection{Image uncertainty} Both the large and the small L-PD network were trained on all the image classes as defined in Table \ref{tab:image_changes} and shown in Fig. \ref{fig:image_changes}. We compared reconstructions of L-PD trained on class 0 to L-PD trained on the same class as the test class. 

\begin{figure}[ht!]
\centering
\includegraphics[width=0.95\linewidth]{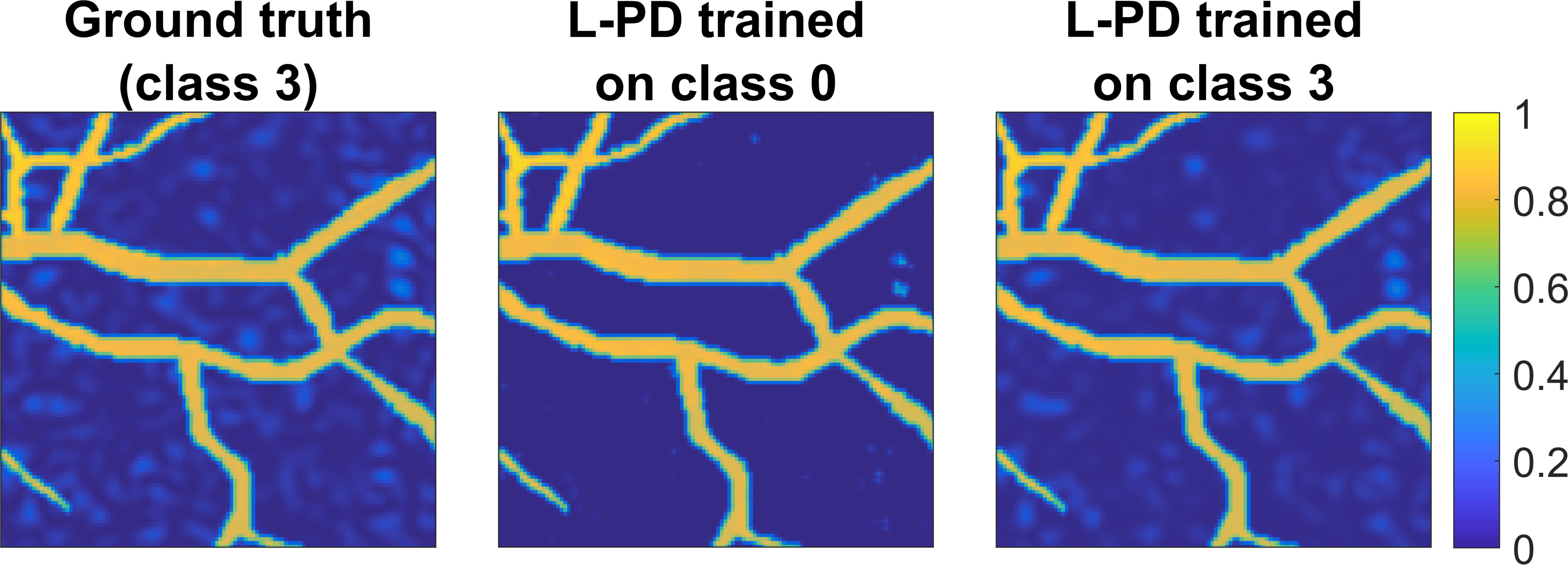}
\caption{Example reconstructions of L-PD trained on different classes, both applied to the same test data (class 3).}
\label{fig:image_changes_result}
\vspace{-3mm}
\end{figure}

One visual comparison is provided in Fig. \ref{fig:image_changes_result}. Here it can be seen that the L-PD network trained on class 0 incorrectly removes most of the background when applied to test data from class 3: only two spots of high background intensity remain. One might argue that this is a positive property, since out-of-focus elements or low-absorbing structures without diagnostic value can be removed. However, spurious high-intensity background elements can be misclassified as a part of the vascular geometry, which is undesired. After training on class 3, the complete background is correctly identified, which makes it less likely that background elements are misclassified as vessels. 

\begin{table*}[ht!]
\vspace{2mm}
\caption {Performance comparison between basic training and training with more variety in image properties.}
\vspace{-3mm}
\label{tab:psnr_image_changes}
\begin{center}
{\rowcolors{2}{white}{gray!25}
\bgroup
\def\arraystretch{1.2}
\begin{tabular}{ l || l | l | l | l | l | l | l | l   } 
 \hline \rowcolor{gray!10} \textbf{\textit{PSNR values for two network sizes}}	&	class $0$:	&	class $1$: 	&	class $2$: 	&	class $3$: 	&	class $4$:	&	class $5$: 	&	class $6$: 	&	class $7$: 	\\
\rowcolor{gray!10}
\textbf{\textit{using different training classes ($\boldsymbol{c}$)}}	&	 nothing	&	diameter	&	contrast	&	 backgr. &	 coverage	&	structure	&	noise level	&	everything	\\ \hline \hline
large network (10 iterations) trained on class $i$	&	\textbf{41.00}	&	39.81	&	\textbf{39.53}	&	\textbf{38.27}	&	\textbf{39.18}	&	\textbf{39.09}	&	\textbf{37.72}	&	\textbf{31.86}	\\
large network (10 iterations) trained on class $0$	&	\textbf{41.00}	&	\textbf{40.78}	&	28.77	&	31.86	&	38.00	&	35.97	&	36.64	&	21.64	\\
small network (5 iterations)$~$ trained on class $i$	&	39.32	&	39.40	&	37.33	&	36.22	&	37.20	&	37.46	&	36.38	&	30.21	\\
small network (5 iterations)$~$ trained on class $0$	&	39.32	&	39.09	&	25.50	&	31.04	&	36.61	&	34.72	&	35.48	&	18.88	\\
 \hline
\end{tabular}
\egroup
}
\end{center}
\vspace{-3mm}
\end{table*}

We compare peak signal-to-noise ratios (PSNR) of the small and large network for the whole test set for all classes. In Table \ref{tab:psnr_image_changes} it can be seen that for almost all image classes, it helps to train on the specific image class instead of the class 0. This is not very surprising, since the specific class is tailored to the test set on which the trained network is applied. It is however interesting to see that the degree of improvement was very different for the different classes: geometrical changes such as diameter (class 1) and coverage (class 4) did not show any significant improvement, while intensity changes such as contrast (class 2) and background (class 3) showed that it was necessary to retrain the network for this specific class. Furthermore it was surprising that retraining for different amounts of noise (class 6) was not really necessary, indicating that the denoising capacity of L-PD seems to be quite robust. {It is not clear to us why for a test set of class 1 the PSNR value is higher when trained on class 0 instead of on class 1; we hypothesise that this is due to the stochasticity of training the neural network in terms of the optimisation.}

It can be seen as well that taking a larger network gives a minor improvement over the small network, but no extreme changes are seen. Finally, it is also apparent that the class that contains all the variety of all the other classes benefits from retraining on the same class, but due to this variety, the PSNR value is lower than for other classes. It is unclear to which extent this could be solved by simply taking a bigger training set. 

\subsubsection{System uncertainty} Next we compare reconstructions of L-PD trained on a fixed setting of 32 detectors to L-PD trained on a random number of detectors, as described in section \ref{sec:system_uncertainty}. {Here only the results for a test set with 64 detectors are shown: the full results are provided in Appendix \ref{app:det}.} In Fig. \ref{fig:system_changes_result}, it can be seen that without training for invariance in detector placement, {there are a lot of undesired local intensity changes in the reconstruction. When training is done with data from a random detector placement, we obtain the desired high-quality reconstruction. This means that L-PD can be applied to data from different number of detectors, without having to perform a time-consuming retraining of the algorithm. }

\begin{figure}[ht!]
\centering
\includegraphics[width=\linewidth]{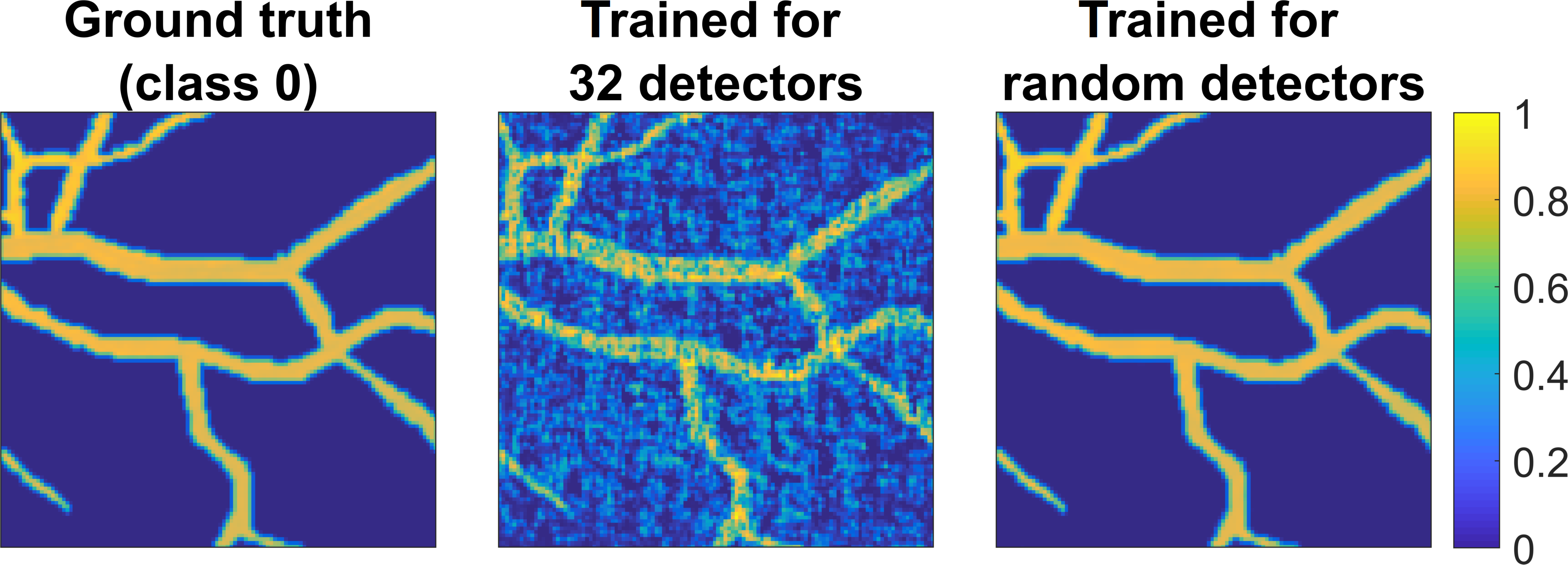}
\caption{Reconstructions of L-PD trained on 32 or a random number of detectors, applied to test data from 64 detectors.}
\label{fig:system_changes_result}
\end{figure}

In Appendix \ref{app:SoS} and Appendix \ref{app:cal} it is shown that a trained L-PD algorithm is robust against heterogeneous sound speed changes and uncertainty in the calibration measurements, respectively.
\vspace{-1.5mm}
\subsection{Joint reconstruction and segmentation}
In this section, the joint reconstruction-segmentation algorithm as explained in section \ref{sec:segmentation} is compared to four different well-known methods.

\begin{figure*}[ht!]
\centering
\includegraphics[width=\linewidth]{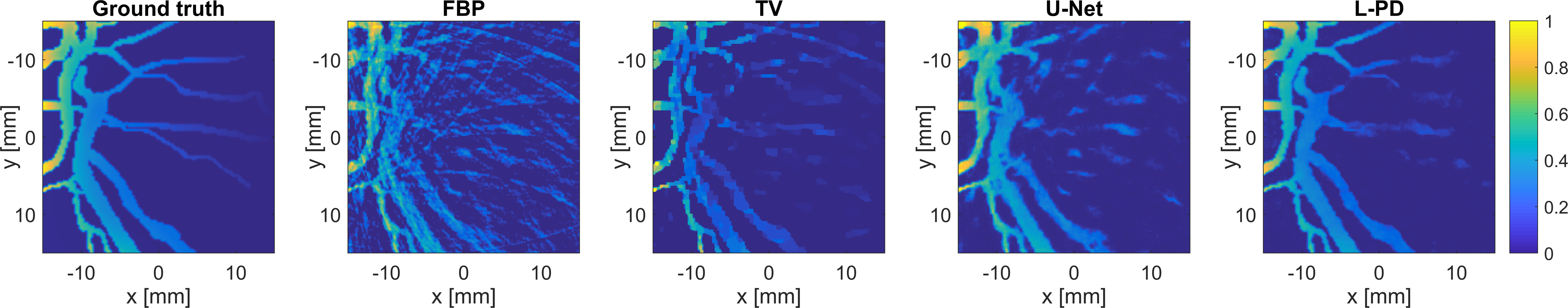}
\caption{Reconstructions of a phantom using a uniform placement of 32 detectors: L-PD gives an improved reconstruction.}
\label{fig:synth_recon}
\end{figure*}

{
\begin{figure*}[ht!]
\centering
\includegraphics[width=0.8\linewidth]{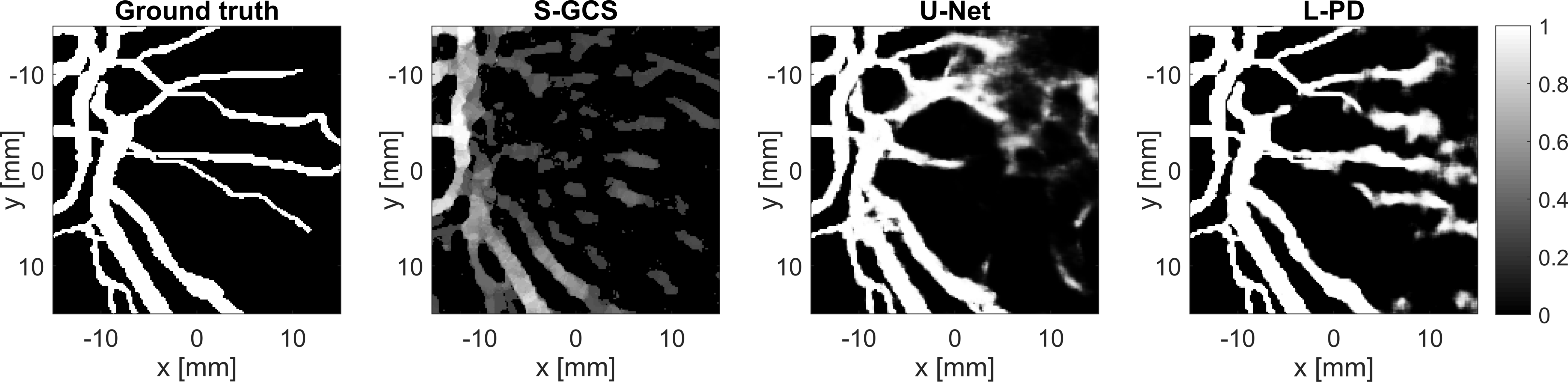}
\caption{Segmentations of the same phantom as in Fig. \ref{fig:synth_recon}. L-PD gives better geometrical information than other methods.}
\label{fig:synth_segm}
\end{figure*}
}

{
\begin{figure*}[ht!]
\centering
\includegraphics[width=0.8\linewidth]{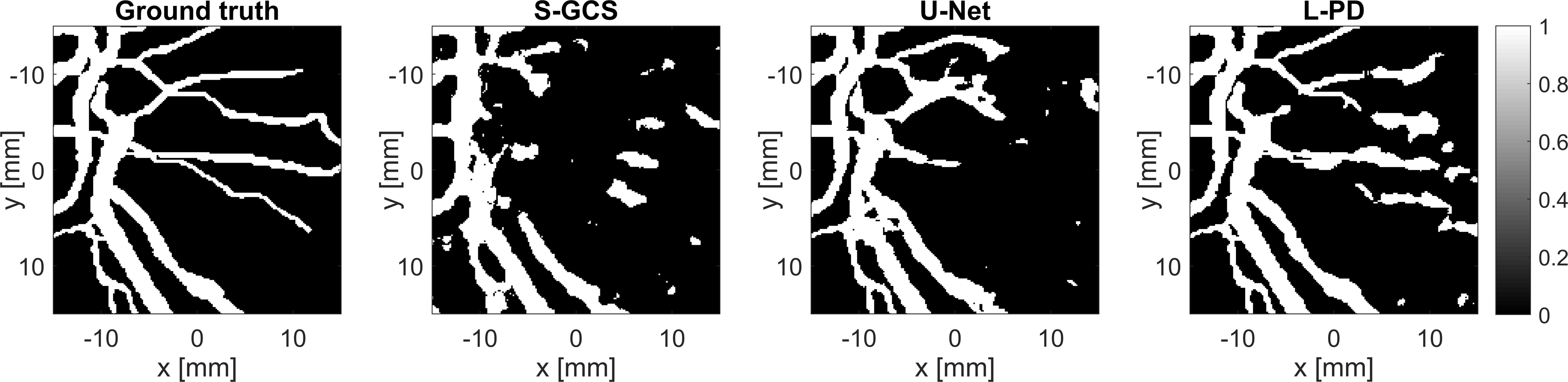}
\caption{Thresholded segmentations from the result shown in Fig. \ref{fig:synth_segm}. L-PD shows connected vascular structures.}
\label{fig:synth_segm_binary}
\end{figure*}
}

In Fig. \ref{fig:synth_recon} a comparison between the different reconstruction methods is shown for synthetic data using a uniform placement of 32 detectors. In this specific phantom, light was modelled to come from the left hand side of the image, resulting in a decaying fluence rate from left to right. It can be seen that TV and U-Net remove part of the background and partly smooth the vascular structure, but only L-PD is able to correctly identify the entire background while retaining most of the vascular structure. 

{The same synthetic data was used for Fig. \ref{fig:synth_segm} and Fig. \ref{fig:synth_segm_binary}, where S-GCS and U-Net are compared to L-PD. Both the non-binary output (Fig. \ref{fig:synth_segm}) and the binary output \ref{fig:synth_segm_binary} indicate the superior performance of the L-PD method. The L-PD method provides the correct vascular geometry, while both S-GCS and U-Net fail in the region where the initial pressure is lower. Furthermore, S-GCS, which is based on a level-set method, connects different structures which are not necessarily connected.} It can be seen that the L-PD segmentation gives somewhat `fuzzy' results in the right part of the image, while it is very sharp in the left part. This indicates that the L-PD segmentation is pixel-wise less reliable when the fluence rate drops. However, it is still clear that four blood vessels are migrating to the right; this possibly important diagnostic information is completely absent in the other segmentation methods.

\begin{figure}[ht!]
\centering
\includegraphics[width=\linewidth]{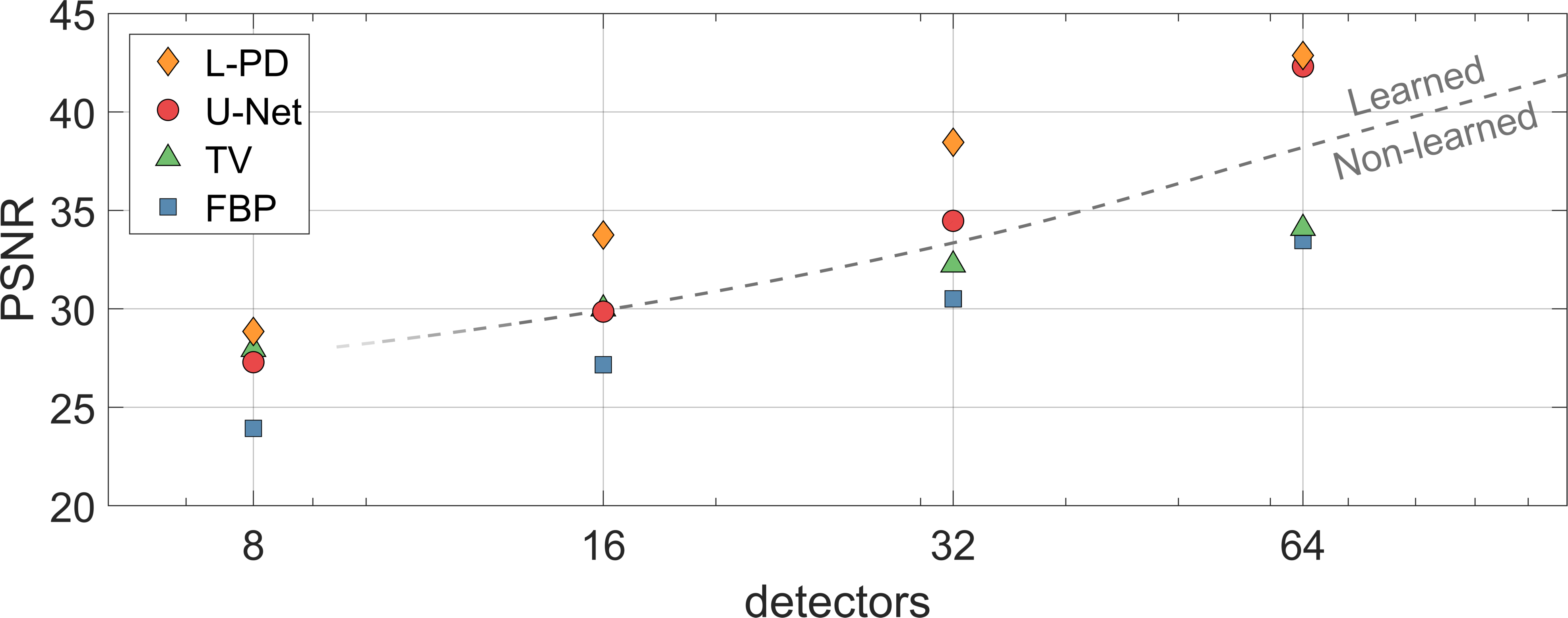}
\caption{Reconstruction quality for four reconstruction methods under uniform compressive sampling: L-PD outperforms other methods in all cases.}
\label{fig:PSNR_recon}
\vspace{-5mm}
\end{figure}

Compressive sampling is a useful way to limit measurement time, thereby minimising influences due to patient movement. We quantitatively show the superior performance of L-PD in both reconstruction and segmentation under uniform compressive samplings. In Fig. \ref{fig:PSNR_recon} it can be seen that both learned reconstruction methods (U-Net and L-PD) give higher quality reconstructions than non-learned methods (FBP and TV). Moreover, it can be seen that L-PD is significantly better than U-Net, especially in case of substantial compressive sampling, meaning a smaller number of detectors. This can be explained by the fact that the FBP-reconstruction with less than 32 detectors sometimes misses important features in the vascular geometry, which cannot be regained with the U-Net approach. With 64 detectors, the FBP-reconstruction does not really miss any important features, so U-Net can be adequately applied as an artefact removal tool.  

{
\begin{figure}[ht!]
\centering
\includegraphics[width=\linewidth]{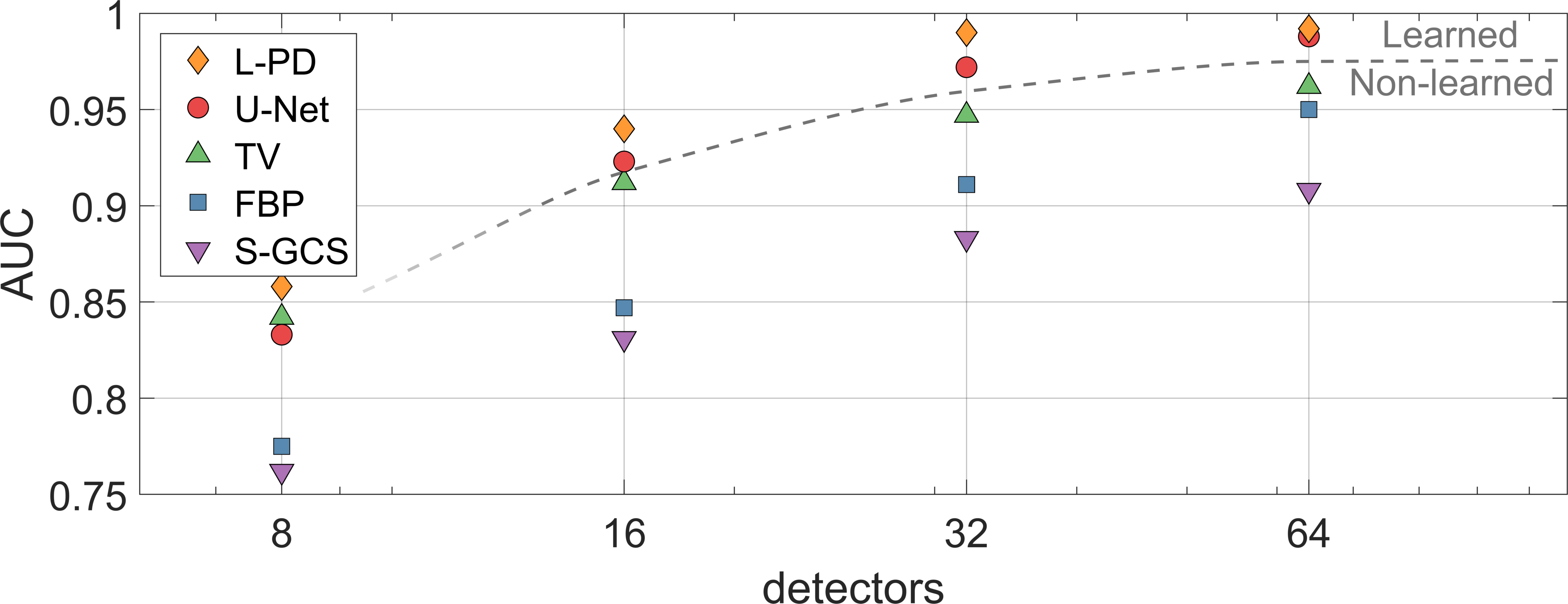}
\caption{Segmentation quality for five segmentation methods under uniform compressive sampling: L-PD outperforms other methods in all cases.}
\label{fig:AUC_recon}
\end{figure}
}

In Fig. \ref{fig:AUC_recon} the segmentation quality is assessed with the area under ROC-curve. The ROC-curve is obtained by comparing the false positive rate to the true positive rate of the thresholded reconstruction for various threshold values. The area under the curve (AUC) is then used as a single value reflecting the segmentation accuracy. The input image for computing the AUC is a non-binary image. In Fig. \ref{fig:AUC_recon} it can be seen that the segmentation quality is very much in line with the reconstruction quality that was shown before: learned methods give more accurate segmentations than non-learned methods. Again, the L-PD method is particularly interesting in case of highly limited number of detectors: it helps to add the acoustic operator in the training, since otherwise previously created artefacts can be difficult to remove.

\subsection{Experimental results}
\begin{figure}[ht!]
\centering
\includegraphics[width=\linewidth]{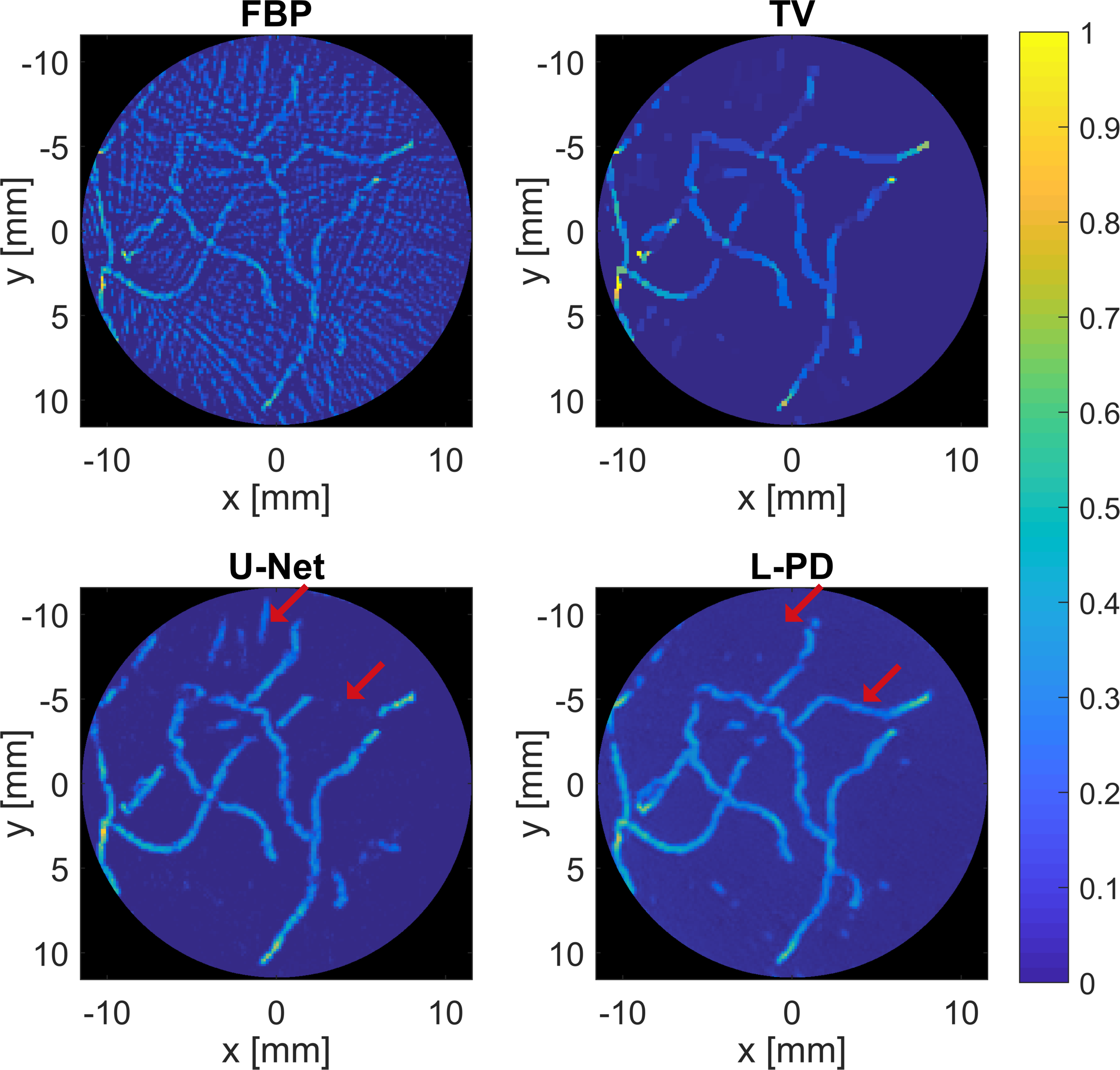}
\caption{Reconstructions of an experimental phantom using a uniform placement of 64 detectors.}
\label{fig:exp_uni_recon}
\end{figure}
\begin{figure}[ht!]
\centering
\includegraphics[width=\linewidth]{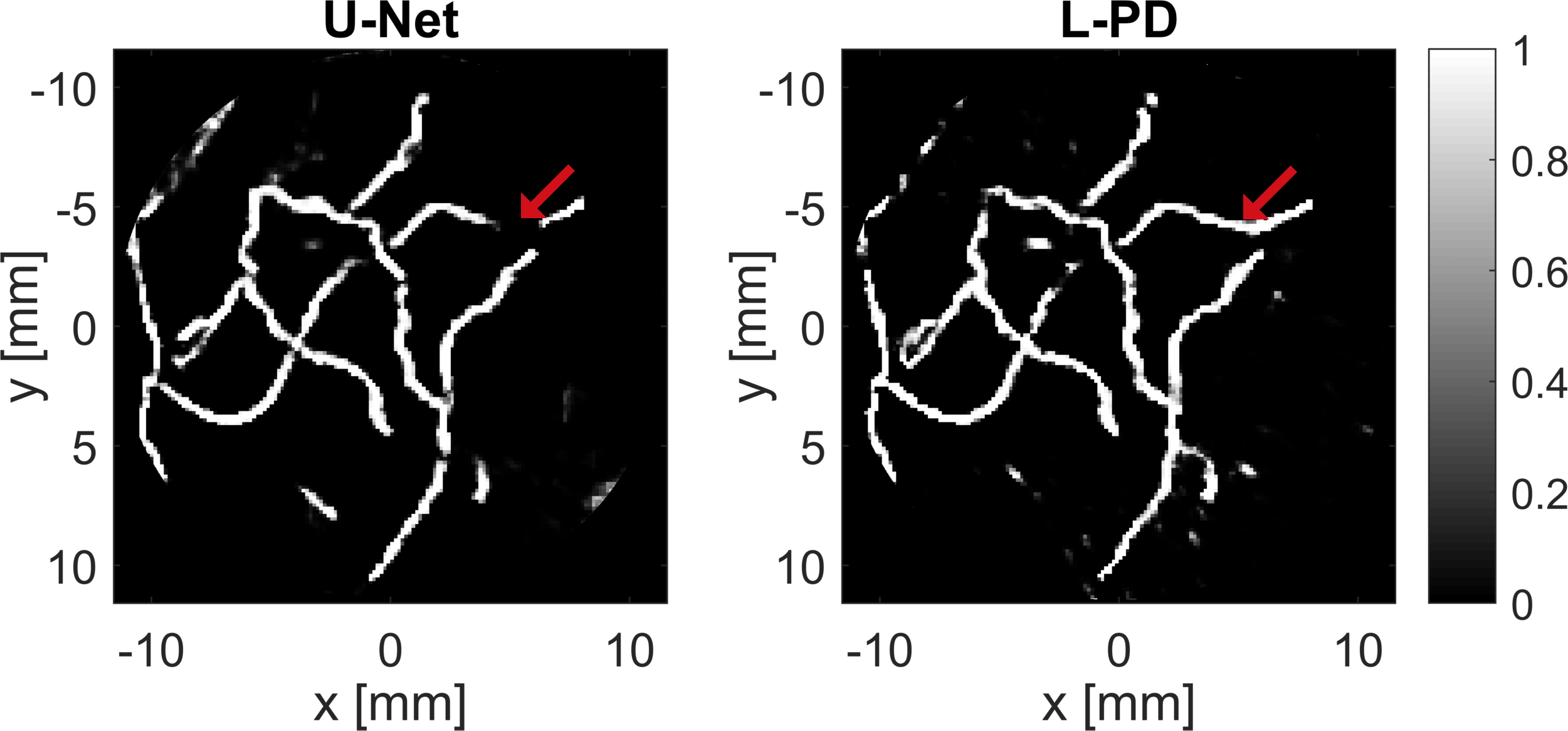}
\caption{Segmentations of an experimental phantom using a uniform placement of 64 detectors.}
\label{fig:exp_uni_segm}\vspace{-3mm}
\end{figure}
All methods have been tested on experimental data obtained by measuring the experimental phantom explained in section \ref{sec:exp_phantom}. Reconstructions and segmentations of the experimental phantom for a uniform sampling of 64 detectors are shown in Fig. \ref{fig:exp_uni_recon} and Fig. \ref{fig:exp_uni_segm}, respectively. Although the sampling of 64 detectors allows for a reasonable reconstruction with either TV or U-Net, the reconstruction with L-PD shows a clearer vascular structure with less artefacts in the background. This is also seen in the segmentation, where a larger part of the vascular geometry is identified. 

\begin{figure}[ht!]
\centering
\includegraphics[width=\linewidth]{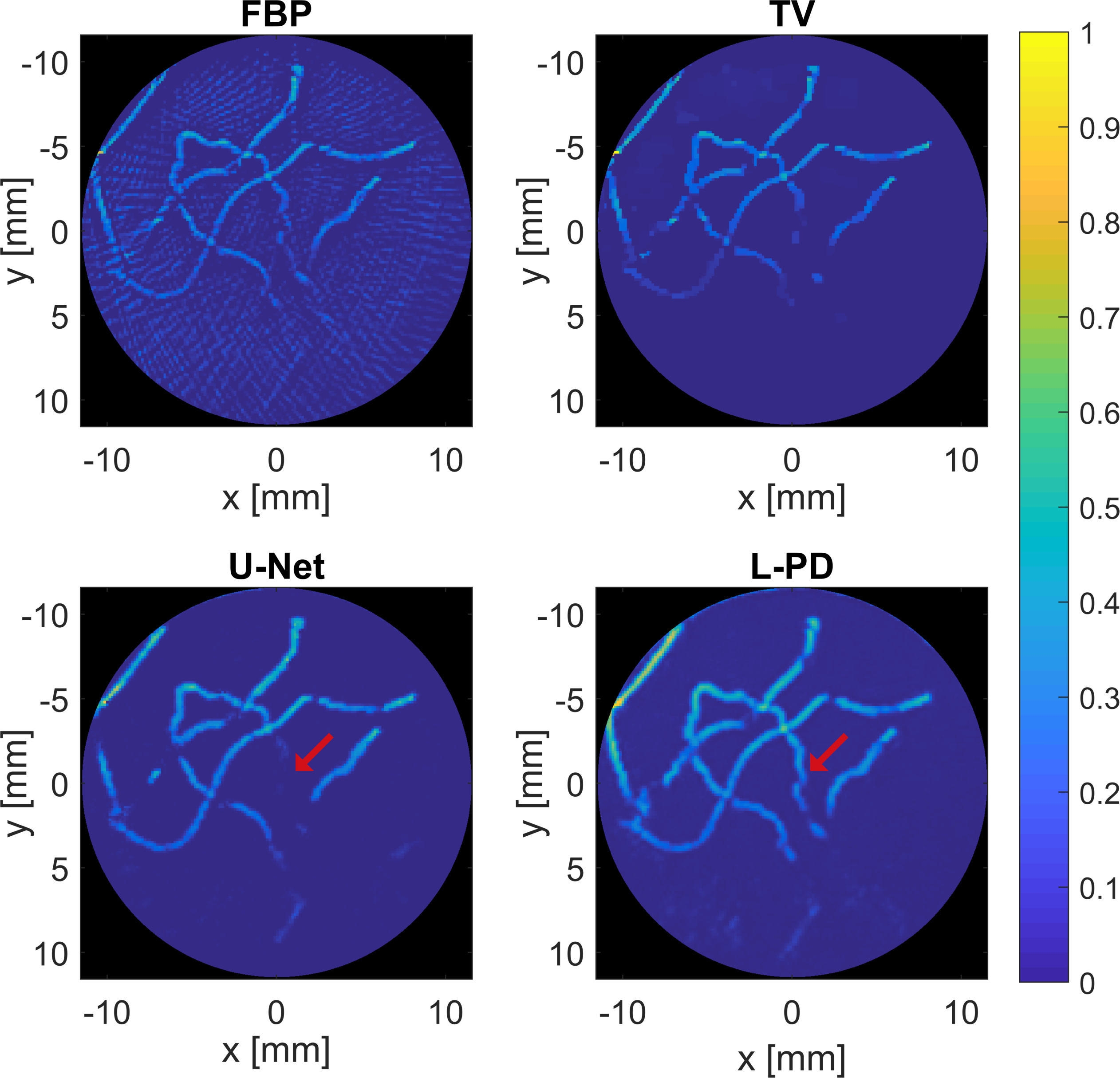}
\caption{Reconstructions of experimental phantom using a limited view placement of 64 detectors at the top of the image.}
\label{fig:exp_nonuni_recon}
\end{figure}

\begin{figure}[ht!]
\centering
\includegraphics[width=\linewidth]{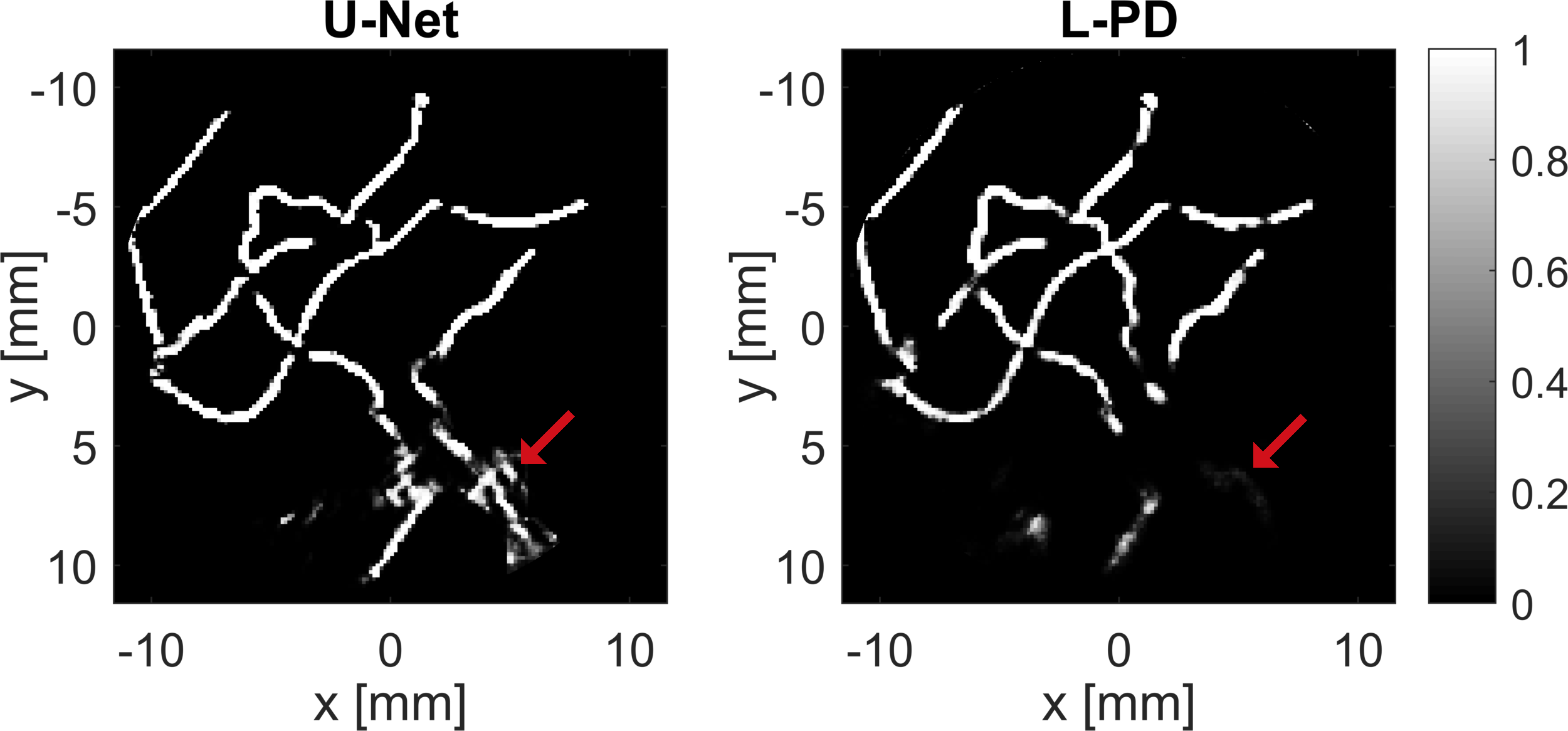}
\caption{Segmentations of experimental phantom using a limited view placement of 64 detectors at the top of the image.}
\label{fig:exp_nonuni_segm}
\end{figure}

In Fig. \ref{fig:exp_nonuni_recon} and Fig. \ref{fig:exp_nonuni_segm}, reconstructions and segmentations for the limited-view sampling of 64 detectors are shown. Recall that in this experimental setup, light deposition and ultrasound detection is at the top half of the phantom. Also here it can be seen that the L-PD approach gives superior reconstructions to the other methods. The L-PD segmentation does not create the spurious parts that are visible in the U-Net segmentation. 

\section{Conclusion and outlook}\label{sec:conclusion}
In this paper we developed a partially learned joint reconstruction and segmentation method for PAT. The sensitivity of the algorithm for reconstruction was investigated. {When suitable training data is considered, the method has shown to be robust to changes in image and system settings and variety in sound speed and calibration measurements. For this, it is assumed that the expected variety can be estimated before training.} Compared to state-of-the-art, our method gives higher quality reconstruction and segmentation results with less computational costs, {both in simulation and experiment}. More specifically, applying our method in a limited-angle scenario results in less streaking artefacts compared to the U-Net approach and can handle smooth intensity gradients. Thus, by improving reconstruction and segmentation quality, photoacoustic tomography can achieve deeper penetration in tissue. The approach can be modified for imaging modalities with different underlying physics or can be extended to other high-level tasks beyond segmentation.

In our experiments only 2D reconstruction and segmentation were considered, while both 2D \cite{Li2018} and 3D systems \cite{Toi2017} are in use. Our method can be applied to 3D without changing the structure of the algorithm. If the 3D photoacoustic operator is computationally too expensive, one might have to change the learning of the algorithm to a step-by-step approach, as was done in \cite{Hauptmann2018}, but this does not change the structure of the convolutional neural network.

One approach to receive the desired robust algorithm is to train on a big variety of input images and system settings. However, when this variety is large, training will take a considerate amount of time. Moreover, it is never possible to train for all possible settings. Therefore, future research could focus on learning a representation of the manifold in which the reconstructions and segmentations should lie. This could be done by combining L-PD with a generative adversarial network \cite{Goodfellow2014} or variational autoencoder \cite{Kingma2013} which learns a low-dimensional representation of the ground truth images. These networks could then learn from a finite number of settings and interpolate on the learned manifold to cover all reasonable inputs and system settings. 

\section*{Acknowledgment}
The authors thank Maura Dantuma for the acquisition of the experimental data. The collaboration project is co-funded by the PPP allowance made available by Health$\sim$Holland, Top Sector Life Sciences \& Health, to stimulate public-private partnerships. SM and CB acknowledge support by the 4TU programme Precision Medicine. SM acknowledges the European Union’s Horizon 2020 research and innovation programme H2020 ICT 2016-2017 under grant agreement No 732411, which is an initiative of the Photonics Public Private Partnership.

\ifCLASSOPTIONcaptionsoff
  \newpage
\fi

\small{
\bibliographystyle{ieeetr} 
\bibliography{refs}
}

\newpage

\appendices

{
\section{Robustness to detector settings: full results}\label{app:det}
In this section the full results of the experiments described in section \ref{sec:system_uncertainty} are shown: we compare reconstructions of L-PD trained on a fixed setting of 32 detectors to L-PD trained on a random number of detectors. In Fig. \ref{fig:detector_settings_0}, it can be seen that the L-PD algorithm trained on a fixed detector setting can not directly be applied to a different detector setting. All reconstructions show significant artefacts, however it is not clear to us why different detector settings give very different artefacts. 

\begin{figure*}[ht!]
\centering
\includegraphics[width=\linewidth]{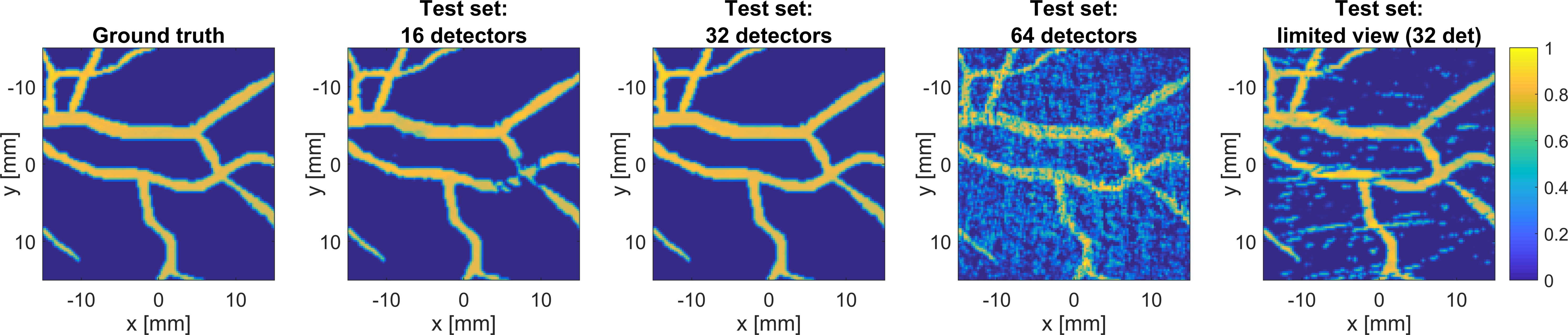}
\caption{L-PD trained on a fixed setting of 32 detectors, tested on different detector settings.}
\label{fig:detector_settings_0}
\end{figure*}

In Fig. \ref{fig:detector_settings_1}, it can be seen that it helps to train on many detector settings, randomly chosen from a possibility of 128 detectors. The reconstruction quality is good for all images.

\begin{figure*}[ht!]
\centering
\includegraphics[width=\linewidth]{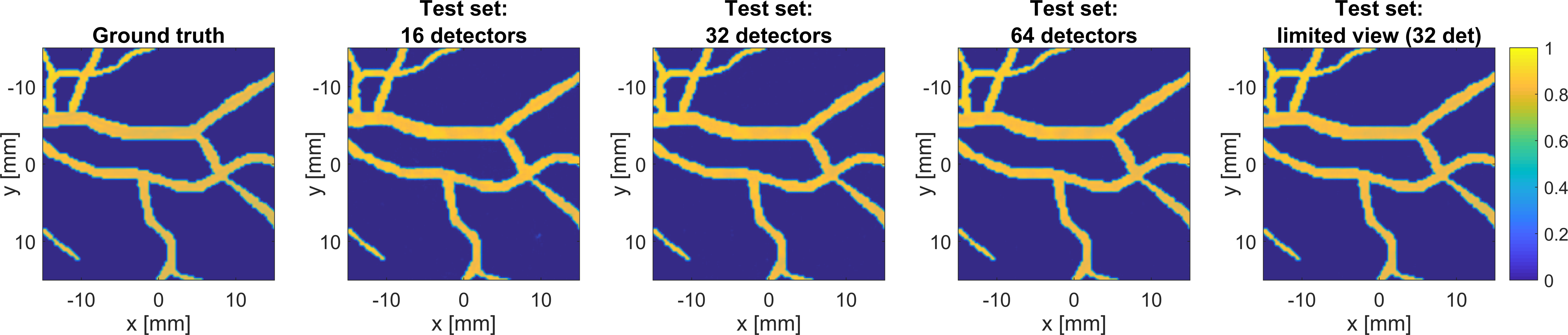}
\caption{L-PD trained on a variable detector setting, tested on different detector settings.}
\label{fig:detector_settings_1}
\end{figure*}

\section{Robustness to varying heterogeneous sound speed}\label{app:SoS}
In this section we analyse the robustness of a trained L-PD algorithm to changes in the sound speed. The algorithm has been trained for a uniform detector setting with 32 detectors, as explained in section \ref{sec:robustness}. Next, the test set has been changed to a scenario where the synthetic phantom has a higher sound speed than the surrounding medium (Fig. \ref{fig:SoS_distribution}). This means that the sound speed is still assumed to be known, but not constant throughout the medium. Derivation of the model with non-constant speed of sound is done by calculating the time-of-flight values, taking refraction of rays into account. The approach is based on solving the Eikonal equation. Details can be found in \cite[section 5.3]{Willemink2010}.

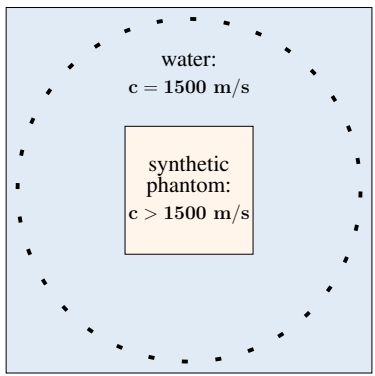
\begin{figure}[ht!]
\centering
\resizebox{0.55\linewidth}{!}{%
\begin{tikzpicture}
\def\centerarc[#1](#2)(#3:#4:#5)
    { \draw[#1] ($(#2)+({#5*cos(#3)},{#5*sin(#3)})$) arc (#3:#4:#5); }

\definecolor{lightbluee}{RGB}{225,235,245}
\definecolor{tissue}{RGB}{255,245,235}

\filldraw[draw=black,fill=lightbluee] (6,4) rectangle (14,-4);
\filldraw[draw=black,fill=tissue] (8.6,1.4) rectangle (11.4,-1.4);

\foreach \i in {0,...,31}
{
	\centerarc[black,line width=2.4](10,0)(11.25*\i-(1.40625+0.703+0.3):11.25*\i+1.40625-(1.40625+0.703)+0.3:3.75)
}

\node (a) at (10,2.5) {\makecell{\Large water:\vspace{1.5mm}\\ \large $\mathbf{c = 1500~m/s}$}};

\node (a) at (10,0) {\makecell{\Large synthetic\\ \Large phantom:\vspace{1.5mm}\\ \large $\mathbf{c > 1500~m/s}$}};
\end{tikzpicture}}
\caption{Sound speed distribution for a uniform detector setting with 32 detectors: the orange region (synthetic phantom) has a higher sound speed than the blue region (water).}
\label{fig:SoS_distribution}
\end{figure}

In Fig. \ref{fig:SoS_results} it can be seen that the reconstruction quality does not become worse when a heterogeneous sound speed is assumed: even if the sound speed is considerably higher within the synthetic phantom (+80 m/s), the reconstruction has the same quality as in the homogeneous sound speed case. 

\begin{figure*}[ht!]
\centering
\includegraphics[width=\linewidth]{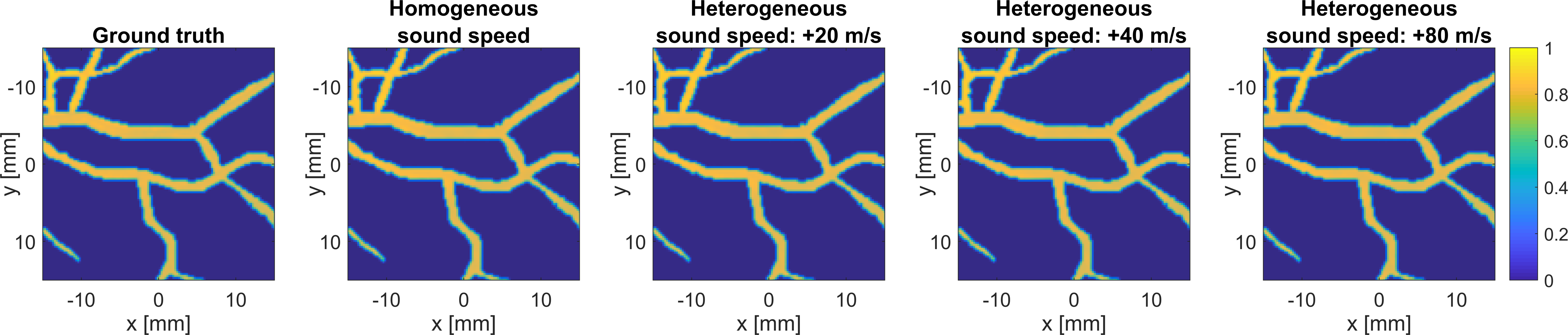}
\caption{L-PD trained for a homogeneous sound speed, tested on data obtained with a heterogeneous sound speed distribution.}
\label{fig:SoS_results}
\end{figure*}

\section{Robustness to different calibrations}\label{app:cal}
Here the robustness of a trained L-PD algorithm to different calibrations is analysed. As explained in section \ref{sec:fw_model}, a deconvolution is applied to the measured data, which gives us the pre-processed data $f$. This pre-processed data is then used as input for the L-PD algorithm. A different calibration measurement $p_\text{cal}$ gives us a different deconvolution, which results in different input data for our L-PD algorithm. 

We have measured four different calibration phantoms at 4 different times. The photoacoustic imaging system was shut down between the separate measurements to cover all system uncertainty that can be expected between any two measurements. The L-PD algorithm has been trained for a uniform detector setting with 32 detectors, as explained in section \ref{sec:robustness}; this was done with calibration 1. Four test sets were created: one used the same calibration 1, the others had a different calibration in their pre-processing. 

\begin{figure*}[ht!]
\centering
\includegraphics[width=\linewidth]{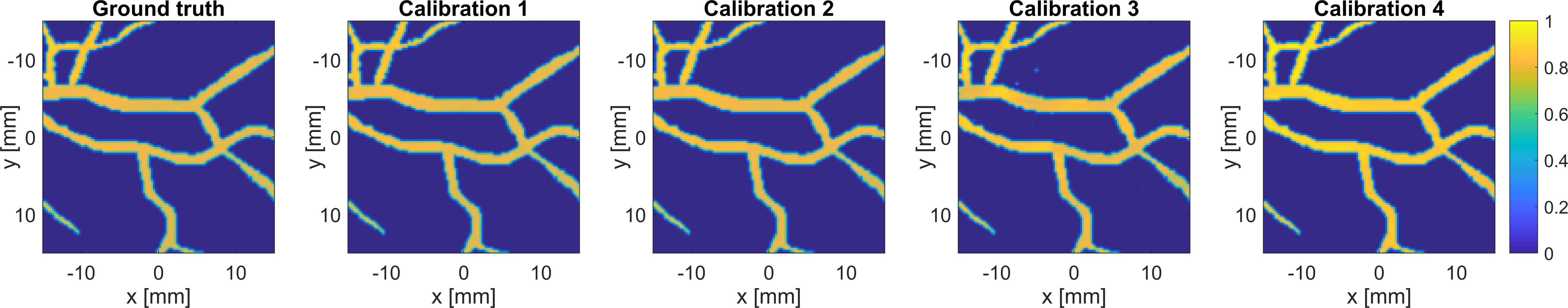}
\caption{L-PD trained on data from calibration 1 and tested on data from four different calibrations.}
\label{fig:Cal_results}
\end{figure*}

In Fig. \ref{fig:Cal_results}, it can be seen that a different calibration has a very minor effect on the reconstruction quality: calibration 2 gives a visually identical result to calibration 1; calibration 3 shows two higher intensity spots in the background; calibration 4 shows a slightly increased overall intensity. The latter can be explained by the fact that the absorbed energy might have been different in the fourth calibration phantom, giving a different scaling when performing a deconvolution.
}
\end{document}